\documentclass[10pt,journal,compsoc]{IEEEtran}

\ifCLASSOPTIONcompsoc
  \usepackage[nocompress]{cite}
\else
  \usepackage{cite}
\fi

\ifCLASSINFOpdf
\else
\fi

\usepackage{amsmath}

\usepackage{amsfonts}
\usepackage{bm}
\usepackage{mathrsfs}
\usepackage{multirow}
 
\usepackage{subfig}
\usepackage{graphicx}
\usepackage{amssymb}
\usepackage[dvipsnames]{xcolor}
\usepackage{booktabs}
\usepackage{ulem}
\usepackage{multirow}

\begin{document}

\title{Deep Coarse-to-fine Dense Light Field Reconstruction with Flexible Sampling and Geometry-aware Fusion}

\author{Jing Jin,~\IEEEmembership{Student Member,~IEEE,}
        Junhui Hou,~\IEEEmembership{Senior Member,~IEEE,}
        Jie Chen,~\IEEEmembership{Member,~IEEE,}
        Huanqiang Zeng,~\IEEEmembership{Senior Member,~IEEE,}
        Sam Kwong,~\IEEEmembership{Fellow,~IEEE,}
        Jingyi Yu

\IEEEcompsocitemizethanks{
\IEEEcompsocthanksitem J. Jin, J. Hou, and S. Kwong are with the Department of Computer Science, City University of Hong Kong, Hong Kong.\protect\\
E-mail: jingjin25-c@my.cityu.edu.hk; \{jh.hou, cssamk\}@cityu.edu.hk
\IEEEcompsocthanksitem J. Chen is with the Department of Computer Science, Hong Kong Baptist University, Hong Kong.\protect\\
E-mail: chenjie@comp.hkbu.edu.hk
\IEEEcompsocthanksitem H. Zeng is with College of Engineering, Huaqiao University, Quanzhou 362021, China.\protect\\
E-mail:zeng0043@hqu.edu.cn
\IEEEcompsocthanksitem J. Yu is with the School of Information Science and Technology, ShanghaiTech University, Shanghai 201210, China.\protect\\
E-mail: yujingyi@shanghaitech.edu.cn
}% <-this % stops an unwanted space
\thanks{This work was supported in part by the Hong Kong RGC under Grant 9048123 (CityU 21211518) and Grant 9042820 (CityU 11219019), in part by the Natural Science Foundation of China under Grant 61871342 and Grant 61871434, and in part by the Basic Research General Program of Shenzhen Municipality under Grant JCYJ20190808183003968. }
\thanks{
Corresponding author: Junhui Hou.}
}

% The paper headers
% \markboth{Manuscript submitted to IEEE Transactions on Pattern Analysis and Machine Intelligence}%
% {Shell \MakeLowercase{\textit{et al.}}: Bare Demo of IEEEtran.cls for Computer Society Journals}

\IEEEtitleabstractindextext{
\begin{abstract}
A densely-sampled light field (LF) is highly desirable in various applications, such as 3-D reconstruction, post-capture refocusing and virtual reality. However, it is costly  to acquire such data.
Although many computational methods have been proposed to reconstruct a densely-sampled  LF from a sparsely-sampled one, they still suffer from either low reconstruction quality, low computational efficiency, or the restriction on the regularity of the sampling pattern. To this end, we propose a novel learning-based method, which accepts sparsely-sampled LFs with irregular structures, and produces densely-sampled LFs with arbitrary angular resolution accurately and efficiently.
We also propose a simple yet effective method for optimizing the sampling pattern. Our proposed method, an end-to-end trainable network,  reconstructs a densely-sampled LF in a coarse-to-fine manner. Specifically, the coarse sub-aperture image (SAI) synthesis module first explores the scene geometry from an unstructured sparsely-sampled LF and leverages it to independently synthesize novel SAIs, in which a confidence-based blending strategy is proposed to fuse the information from different input SAIs, giving an intermediate densely-sampled LF.  Then, the efficient LF refinement module learns the angular  relationship within the intermediate result to recover the LF parallax structure. Comprehensive experimental evaluations demonstrate the superiority of our method on both real-world and synthetic LF images when compared with state-of-the-art methods.
In addition, we illustrate the benefits and advantages of the proposed approach when applied in various LF-based applications, including image-based rendering and depth estimation enhancement.
The code is available at https://github.com/jingjin25/LFASR-FS-GAF.
\end{abstract}

\begin{IEEEkeywords}
Light field, deep learning, depth estimation, super resolution, compression, image-based rendering.
\end{IEEEkeywords}}

\maketitle

\IEEEdisplaynontitleabstractindextext

\IEEEpeerreviewmaketitle

\IEEEraisesectionheading{\section{Introduction}\label{sec:introduction}}

\IEEEPARstart{T}{he} light field (LF) is a high-dimensional function describing light rays through every point traveling in every direction in the free space \cite{lf1996rendering,lf1996lumigraph}.
This function is initially introduced for LF rendering, which is an attractive method for generating novel views from a given set of pre-acquired views. Instead of the traditional image-based rendering (IBR) methods, LF rendering treats the captured images as samples of the LF function, and the novel views can be generated by re-sampling a slice from the function in real-time, during which no geometry information is required. To avoid ghosting effects, the LF is required to be densely sampled
 \cite{chai2000plenopticsampling}. 
Densely-sampled  LFs including sufficient information will also facilitate  a wide range of applications, such as
accurate depth inference \cite{lfdepth2014wanner,lfdepth2018chen}, 3-D scene reconstruction \cite{lfapp2013scene} and post-capture refocusing \cite{lfapp2014refocus}.
In addition, with the rapid development of virtual reality technology, a densely-sampled LF becomes vital as it provides smooth angular parallax shift as well as natural focus details, which are important for a satisfying immersive viewing experience \cite{LFVR2017yu,LFVR2015display,LFVR2018system}.

\par
The densely-sampled LF is highly desirable but raises great challenges for the acquisition.
For example, LF images with high angular resolution can be captured using a camera array \cite{lf2005array} for simultaneous sampling from different viewpoints or computer-controlled gantry \cite{lfgantry} for time-sequential sampling at different positions.
However, the former is expensive and bulky, and the latter is limited to static scenes.
The commercialization of hand-held LF cameras such as Lytro \cite{lytro} and Raytrix \cite{raytrix} makes it convenient to acquire LF images. 
These cameras are cheaper and portable by encoding 4-D LF data into a single 2-D sensor.
However, due to limited sensor resolution, a trade-off between spatial and angular resolution exists.

\par
Instead of relying on the development of hardware, many computational methods have been proposed for reconstructing a densely-sampled LF from a sparse one, which can be realized with low cost commercial devices.
Previous works \cite{lfdepth2015wang,lfrec2012gmm,lfrec2013compressive,lfrec2014sparsity,lfrec2015phase,lfrec2018shearlet}
either estimate disparity maps as auxiliary information, 
or use specific priors such as sparsity in transformation domain for dense reconstruction.
With recent development of deep learning solutions for visual modeling, some learning-based methods \cite{lfrec2015yoon,lfrec2016kalantari,lfrec2018wu:blur,lfrec2019wu:shear} have been proposed.
However, most of the existing methods require the input sub-aperture images (SAIs) to be sampled with a specific or regular pattern, which raises difficulties for practical acquisition.
Moreover, since the scene geometry is inexplicitly and insufficiently modeled in these methods, the aliasing problem becomes serious in the reconstructed images when the input LF is extremely under-sampled, i.e.  the samples have large disparities.

\par
As a preliminary work \cite{lfrec2018Yeung}, we proposed a learning-based model for densely-sampled LF reconstruction.
The reconstruction of all novel SAIs are performed in one forward pass during which the intrinsic LF structural information among them is fully explored. See more details in Section \ref{sec:related_lfrec}.
Although this method can produce impressive and state-of-the-art results on extensive real-world images captured by the Lytro Illum camera, the performance degradation caused by sparse sampling and the problem of non-flexibility still exits.
In this paper, built upon \cite{lfrec2018Yeung}, we provide a few distinguishable improvements, enabling flexible and accurate reconstruction of a densely-sampled  LF from sparse sampling.
We inherit the coarse-to-fine framework in \cite{lfrec2018Yeung}. That is, the proposed model consists of two modules, namely the coarse SAI synthesis and the efficient LF refinement.
Specifically, the coarse SAI synthesis module independently synthesizes novel SAIs using geometry-based warping, where we take the sampling with large disparities and arbitrary patterns into consideration.
We also propose a novel confidence-based strategy for handling the occluded regions when blending the warped images from different viewpoints.
We further refine the coarse results by exploiting all the intermediate SAIs with efficient pseudo 4-D filters. Such a refinement module is capable of improving the reconstruction quality by utilizing the  intrinsic LF parallax structure.

In summary, the main contributions of this paper are as follows: 
\begin{itemize}
    \item we propose an end-to-end learning-based method for the reconstruction of densely-sampled  LFs from sparsely-sampled  LFs. Our method maintains high reconstruction quality when the sampling disparity increases, and improves the generality by enabling flexible input positions as well as flexible output angular resolution. We also propose effective strategies for handling occlusions and preserving the LF parallax structure;
    \item we investigate how the sampling pattern affects the reconstruction quality, and propose a simple yet effective method for optimizing the sampling pattern;
    \item we design various and extensive experiments to evaluate and analyze our method as well as those under comparison comprehensively; and
    \item we demonstrate and discuss the benefits of the proposed approach to  LF-based downstream applications.
\end{itemize}
\par
The rest of this paper is organized as follows. 
Sec. \ref{sec:related} comprehensively reviews existing methods for view synthesis and densely-sampled LF reconstruction. 
Sec. \ref{sec:approach} presents the proposed approach and  investigates the optimization for sampling patterns.
In Sec. \ref{sec:experiment}, extensive experiments are carried out to evaluate the performance of the proposed approach.
The benefits of the proposed approach to practical LF-based applications are validated and discussed in Sec. \ref{sec:application}.
Finally, Sec. \ref{sec:conclusion} concludes this paper.

\section{Related Work}
\label{sec:related}

\subsection{View Synthesis}
View synthesis, taking one or more views as inputs to render novel views, is a long-standing problem in the field of computer graphics and computer vision.
Most algorithms leverage the scene geometry information for view synthesis, that is, to extract/learn the global/local geometry  from the input viewpoints and use the resulting geometry information to warp the input views, followed by blending for novel view rendering \cite{viewsyn1993viewinter,viewsyn1995IBRsynstem}. 
However, the forward warping operation typically leads to a hole-filling problem in occlusion areas.
Flynn \textit{et al.} \cite{viewsyn2016flynn} proposed to project input views to a set of depth planes and learn the weights to average the color of each plane.
This method needs to learn specific geometry for different target viewpoints.
To overcome this shortage, some methods based on 3-D scene representation were proposed. Penner \textit{et al.} \cite{viewsyn2017soft} presented a soft 3-D representation by preserving depth uncertainty. Tulsiani \textit{et al.} \cite{viewsyn2018layer} modeled the 3-D structure of the scene by
learning to predict a layer-based representation, which represents multiple ordered depths per pixel along with color values.
Zhou \textit{et al.} \cite{viewsyn2018magnification} proposed to use multi-plane images where each plane encodes color and transparency maps. Through these methods, novel views at varying positions can be rendered by simply forward projecting their corresponding representations. Besides, many methods aim at reconstructing 3-D scenes and synthesizing novel views from a single image (e.g., \cite{viewsyn2016appearance,viewsyn2017consistency,viewsyn2017transform,viewsyn2016multi}). However, these methods are still limited over simple and non-photorealistic synthetic objects.

    \begin{figure*}[!t]
    \centering
    \includegraphics[width=\linewidth]{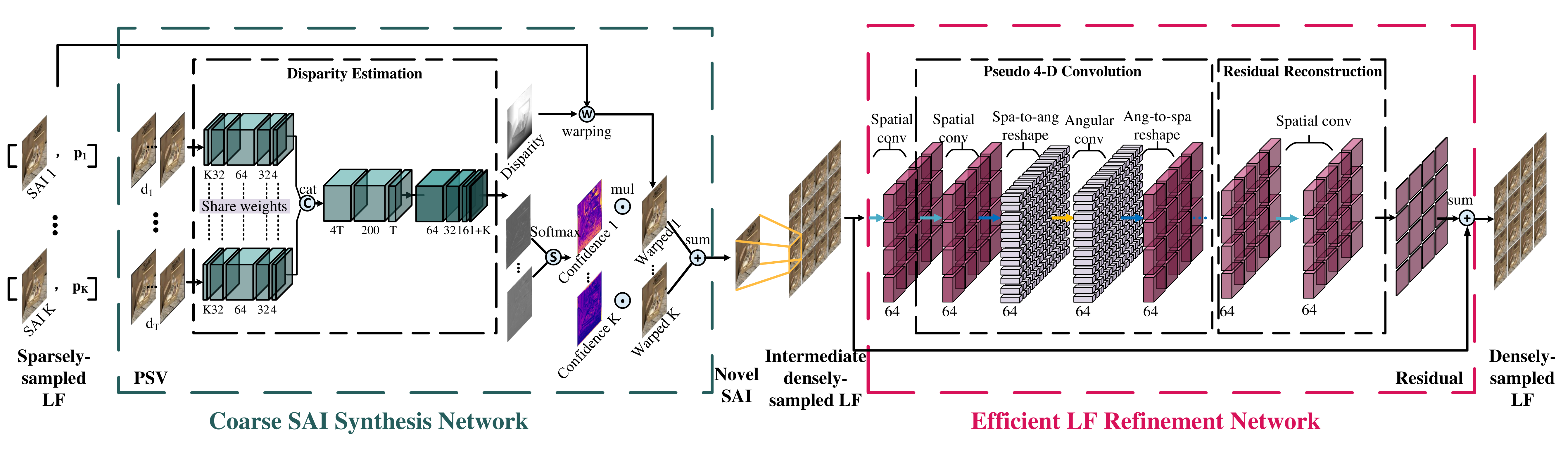}
    \caption{The flowchart of the proposed method for reconstructing a densely-sampled LF with $M\times N$ SAIs from a sparsely- and arbitrarily-sampled LF with $K$ SAIs. Our proposed model consists of two phases, i.e., the coarse SAI synthesis and the efficient LF refinement.}
    \label{fig_workflow}
    \end{figure*}

\subsection{LF Reconstruction}
\label{sec:related_lfrec}

LF rendering needs densely-sampled LFs as inputs.
In what follows, we only focus on the methods that reconstruct a densely-sampled  LF from a  sparsely-sampled one.
Available solutions can be roughly classified to two categorizes: non-learning based methods and learning based methods.

\textbf{Non-learning based methods.}
Many traditional solutions that are originally adopted for natural image processing, such as Gaussian model and sparse representation, have been explored for LF processing tasks. Among them, Mitra \textit{et al.} \cite{lfrec2012gmm} modeled the LF patches using a Gaussian mixture model to address many LF processing tasks. 
Although it can achieve promising results to a certain extent, it is not robust against noise. Shi \textit{et al.} \cite{lfrec2014sparsity} explored sparsity in the continuous Fourier domain to reconstruct densely-sampled  LFs from a small set of samples. Vagharshakyan \textit{et al.} \cite{lfrec2018shearlet} proposed an approach using the sparse representation of epipolar-plane images (EPIs) in the shearlet transform domain. These methods require the sparsely-sampled LF to be sampled in a regular grid. Moreover, some methods explore the compressive LF photography. Marwah \textit{et al.} \cite{lfrec2013compressive} proposed a compressive LF camera architecture which allows LF reconstruction based on overcomplete dictionaries. To reduce the computational cost for dictionary learning, Kamal \textit{et al.} \cite{lfrec2016tensor} exploited a joint tensor low-rank and sparse prior for compressive reconstruction. These methods were specifically designed for coded LF acquisition.

Many works on LF reconstruction leverage explicit depth information for LF reconstruction.
Zhang \textit{et al.} \cite{lfrec2015phase} proposed a depth-assisted phase-based synthesis strategy for a micro-baseline stereo pair. 
% However, this approach  needs iterations for optimization.
Patch-based synthesis methods were presented by Zhang \textit{et al.} \cite{lfrec2017patch}, in which the center SAI is decomposed into different depth layers and  LF editing is performed on all layers. 
However, this method has limited performance for view synthesis, especially for complex scenes.
Some works were developed based on the idea of warping given SAIs to novel SAIs guided by an estimated disparity map. 
Wanner and Goldluecke \cite{lfdepth2014wanner} formulated the SAI synthesis problem as an energy minimization problem with a total variation prior, where the disparity map is obtained through global optimization with a structure tensor computed on the 2-D EPI slices. 
This approach considers disparity estimation as a separate step from view synthesis, which makes the reconstruction quality heavily depend on the accuracy of the estimated disparity maps. Although subsequent research \cite{lfdepth2015jeon,lfdepth2015wang,lfdepth2018chen} has shown significantly better disparity estimations, ghosting and tearing effects are still presented.

\par 
\textbf{Learning-based methods.}
With the great success of deep convolutional neural networks in the field of image processing \cite{sisr2016srcnn,sisr2016espcn,sisr2016vdsr,sisr2017lapsrn}, many learning-based methods have been proposed for densely-sampled LF reconstruction.
Yoon \textit{et al.} \cite{lfrec2015yoon} jointly super-resolved the LF image in both spatial and angular domain using a network that closely resembles the model proposed in \cite{sisr2014srcnn}. Their approach is limited to scale 2 angular super-resolution and cannot flexibly adapt to  sparsely-sampled LF inputs. 
Following the idea of single image super-resolution, Wu \textit{et al.} \cite{lfrec2017wu:blur,lfrec2018wu:blur} proposed an LF reconstruction method which focuses on recovering the high frequency details of the bicubic up-sampled EPIs. In these methods, a blur-deblur scheme was proposed to address the information asymmetry problem caused by sparse angular sampling. 
Based on the observation that an EPI shows clear structure when sheared with the disparity value,
Wu \textit{et al.} \cite{lfrec2019wu:shear} proposed to fuse a set of sheared EPIs for LF reconstruction.
Wang \textit{et al.} \cite{lfrec2018p4d} also proposed a method based on EPIs, which applies 3-D convolutional layers to recover the details on horizontal and vertical EPIs sequentially.
However, since each EPI is a 2-D slice of the 4-D LF, the accessible spatial and angular information of these EPI-based models is severely restricted.
Moreover, for these models,
 novel SAIs must be synthesized horizontally or vertically in 2-D angular domain, resulting in accumulated errors.
Yeung \textit{et al.} \cite{lfrec2018Yeung} proposed an end-to-end network for densely-sampled LF reconstruction. By exploring  the  relationships between SAIs with pseudo 4-D filters, this method achieves state-of-the-art performance over a large number of real-world scenes captured by the Lytro camera.

\par
In addition, depth information is also utilized in some learning-based methods for LF reconstruction.
Srinivasan \textit{et al.} \cite{lfrec2017single} proposed to synthesize a 4-D LF image from a 2-D RGB image based on estimated 4-D ray depth.
However, this method requires a large training dataset and only works on simple scenes since the information contained in single 2-D images is extremely limited.
Kalantari \textit{et al.} \cite{lfrec2016kalantari} proposed to synthesize novel SAIs with two sequential networks that perform depth estimation and color prediction successively. Although this method achieves good performance on LF images captured by the Lytro camera, the depth estimation and color prediction module are implemented in a straightforward manner, which leaves room for improvement.
Jin \textit{et al.} \cite{lfrec2020aaai} also proposed to make use of the geometry information to handle LF images with large disparities.

\section{The Proposed Approach}
\label{sec:approach}
\subsection{4-D LF and Problem Formulation}
A 4-D LF can be represented with the two-plane parameterization structure, which uniquely describes the propagation direction of a light ray via two points from two parallel planes, i.e.,
the angular plane $(u,v)$ and the spatial plane $(x,y)$. Let $\mathcal{I}\in\mathbb{R}^{W\times H\times M\times N}$ denote a densely-sampled  LF containing $M\times N$ SAIs of spatial dimension $W\times H$, which are sampled on the angular plane with a regular 2-D grid of size $M\times N$.
Let $\mathcal{U}$ be the set of  2-D angular coordinates of the SAIs  in $\mathcal{I}$, 
i.e. $\mathcal{U} = \left \{\bm{u} | \bm{u}=(u,v), 1\leq u\leq M, 1\leq v \leq N \right\}$.
The SAI at $\bm{u}$ is denoted as $\bm{I}_{\bm{u}} \in \mathbb{R}^{W\times H}$.
Let $\mathcal{I}_s$ denote a sparsely-sampled  LF with $K$ SAIs, $\mathcal{P}$ be the set of the 2-D angular coordinates of the SAIs in $\mathcal{I}_s$, i.e., $\mathcal{P}=\left\{ \bm{p}_k |\bm{p}_k=(u,v), 1\leq k \leq K \right\}$, and $\bm{I}_{\bm{p}_k}$ be an SAI in $\mathcal{I}_s$ located at $\bm{p}_k$.
Moreover, the SAIs of a sparsely-sampled  LF are assumed to be arbitrarily sampled from a certain densely-sampled  LF, i.e., $\mathcal{P}\subset\mathcal{U}$ and $K\ll MN$.
The unsampled SAIs, which belong to $\mathcal{I}$ but do not appear in $\mathcal{I}_s$ are denoted by $\mathcal{I}_{\overline{s}}=\{\bm{I}_{\bm{q}_l}|\bm{q}_l\in \mathcal{Q} = \mathcal{U} \setminus \mathcal{P}, 1\leq l\leq MN-K\}$ with the operator $\setminus$ returing the difference between two sets.

\par
Our goal is to learn  $\widehat{\mathcal{I}}_{\overline{s}}$ as close to $\mathcal{I}_{\overline{s}}$ as possible based on $\mathcal{I}_s$ such that a densely-sampled  LF denoted by $\widehat{\mathcal{I}}\in\mathbb{R}^{W\times H\times M\times N}$ can be reconstructed, together with $\mathcal{I}_s$.
This problem can be implicitly formulated as:
\begin{equation}
  \widehat{\mathcal{I}} =  \mathcal{I}_s \bigcup \widehat{\mathcal{I}}_{\overline{s}} = 
f\left(\mathcal{I}_s,\mathcal{P},\mathcal{Q}\right),
\end{equation}
where $f$ denotes the mapping function to be learnt, and $\bigcup$ is the operator to combine two sets.

\subsection{Overview of the Proposed Method}
SAIs in $\mathcal{I}$ are correlated to each other, which reveals the LF parallax structure.
Specifically,
under the Lambertian assumption and in the absence of occlusions, the  relationship between SAIs of $\mathcal{I}$ can be expressed as
\begin{equation}
   \bm{I}_{\bm{u}} (\bm{x}) = \bm{I}_{\bm{u}+\Delta\bm{u}}(\bm{x}+d\Delta\bm{u}),
\end{equation}
where $\bm{x}=(x,y)$ is the spatial coordinates, and $d$ is the disparity at the pixel $\bm{I}_{\bm{u}} (\bm{x})$.
Being aware of this unique characteristic as well as the great success of deep learning, we propose a learning-based approach to explore the LF parallax structure for densely-sampled  LF reconstruction, i.e., constructing a deep network to learn $f$, as shown in Fig. \ref{fig_workflow}.
Our approach consists of two modules, namely the coarse SAI synthesis network $f^c(\cdot)$ and  the LF refinement network $f^r(\cdot)$, which predicts $\widehat{\mathcal{I}}$ in a coarse-to-fine manner.
To be specific, by explicitly learning the scene geometry from input SAIs, the coarse SAI synthesis network individually generates novel SAIs, giving an intermediate densely-sampled  LF denoted as $\widetilde{\mathcal{I}}$:
\begin{equation}
   \widetilde{\mathcal{I}} = \mathcal{I}_s \bigcup \widetilde{\mathcal{I}}_{\overline{s}} = f^c(\mathcal{I}_s,\mathcal{P},\mathcal{Q}).
\end{equation}
The independent synthesis of the novel SAIs greatly saves computational time and memory usage during testing stage.
Then, the efficient refinement network learns residuals for $\widetilde{\mathcal{I}}$ by exploring the complementary information between the SAIs to recover the LF parallax structure, leading to the final output:
\begin{equation}
\label{eq:refine}
   \widehat{\mathcal{I}} = \widetilde{\mathcal{I}}  + f^r\left( \widetilde{\mathcal{I}} \right).
\end{equation}

\par
By characterizing the sparsely- and densely-sampled  LFs,
our approach improves the flexibility and accuracy of the reconstruction of a densely-sampled LF.
Specifically, our approach has the following characteristics:
\begin{itemize}
\item it overcomes the aliasing problem caused by sparse sampling, making it possible for sparsely-sampled  LFs with different angular sampling rate as inputs;
\item it enables SAIs with arbitrary angular sampling patterns to be used as inputs, which brings more flexibility for the densely-sampled LF reconstruction. Moreover, we further investigated to optimize the sampling patterns for improving reconstruction quality;
\item beyond the early mentioned goal, our method can produce densely-sampled  LFs with user-defined angular resolution, making it more flexible for densely-sampled LF reconstruction in various scenes; and
\item  it is able to accurately recover the valuable LF parallax structure, which is crucial for various applications based on a densely-sampled LF.
\end{itemize}
In the following, the details of the proposed approach are presented step-by-step.

\subsection{Coarse SAI Synthesis}
This module aims at independently synthesizing intermediate novel SAIs denoted by  $\widetilde{\mathcal{I}}_{\overline{s}} = \left\{\widetilde{\bm{I}}_{\bm{q}_l}\right\}$,
which is formulated as
\begin{equation}
    \widetilde{\bm{I}}_{\bm{q}_l} = f^c\left( \mathcal{I}_s, \mathcal{P}, \bm{q}_l  \right).
\end{equation}
To handle the inputs with large disparities, we utilize the geometry information explicitly for novel SAI synthesis. That is, we learn the disparity map at $\bm{q}_l$ from $\mathcal{I}_s$ and synthesize the target SAI via backward warping.
To deal with the challenge posed by the irregular sampling patterns, we construct the disparity estimation network by learning correspondence from the plane-sweep volumes (PSVs) \cite{viewsyn1996psv}.
We also propose a new strategy for blending the warped images, which is able to alleviate the artifacts around occlusion boundaries caused by warping.
To this end, this module consists of three steps: PSV construction, disparity estimation, warping and blending.

\par
\textbf{PSV construction}.
A naive way of disparity estimation is via directly extracting features from $\mathcal{I}_s$ using sequential convolutional layers.
However, for randomly-sampled SAI inputs, i.e. the angular position set $\mathcal{P}$ always varies, it is difficult to properly provide the network with indicators w.r.t the sampling and target positions, making the prediction unreliable (see results in Fig. \ref{fig_disp_inter}).
Instead, we use PSVs for disparity estimation.
A PSV with respect to a target position $\bm{q}_l$ is constructed by backward warping, i.e., reprojecting $\mathcal{I}_s=\left\{ \bm{I}_{\bm{p}_k}\right\}$ with respect to a set of disparity planes $\{d\}$, 
resulting in a set of warped images $\mathcal{V}= \left\{\bm{V}^k_d\right\}$:
\begin{equation}
    \bm{V}^k_d (\bm{x}) = \bm{I}_{\bm{p}_k}(\bm{x}+d(\bm{q}_l-\bm{p}_k)).
\end{equation}
In this way, the arbitrary sampling positions of input SAIs as well as the target position for synthesis are encoded into the PSVs during its construction.

\par
The disparity inference from a PSV is based on principles of photo-consistency. However, in occlusion areas or non-Lambertian surfaces, the  relationships between the matching patches of different SAIs are complicated.
We propose to feed the whole PSV into the disparity estimation network, which is different from the way adopted in \cite{lfrec2016kalantari}, where simple hand-craft features such as mean and standard deviation of the PSV across disparity planes are used.
With the convolutional network's  powerful ability in learning the representation, 
we are able to accurately estimate the disparity maps at challenging regions with the rich information provided by the PSVs.

\par
\textbf{Disparity estimation}.
The disparity estimation network is designed to predict a disparity map $\bm{D}_{\bm{q}_l}$ at the target position $\bm{q}_l$  based on $\mathcal{V}$.
The network consists of a cost calculator to learn the matching cost for each disparity plane, and an estimator to predict the disparity value.

\par
For cost calculator, several convolutional layers are applied to per disparity plane using shared weights.
For a typical disparity plane $d^{\ast}$,
features measuring the similarity and diversity between images warped from different input SAIs are extracted  from $ \left\{\bm{V}^k_{d^{\ast}}\right\}$.
We use kernel size $5\times 5$ to obtain a relatively large receptive field and 
set the number of channels in the final layer as $4$ in the cost calculator.
For the disparity estimator, all features from each disparity plane are concatenated together.
Then sequential convolutional layers are used to predict the disparity value.
Instead of selecting the disparity value with a minimum cost from the predefined disparity set, we let the network learn the disparity value, so that the number of the predefined disparity plane, as well as the width of the network (i.e., the channel number), can be reduced.
The number of channels in the hidden layers of the estimator is set to $200$ at the front layer, and then gradually decreased from $200$ to $64$, $32$, $16$ and $1$ to output a disparity map $\bm{D}_{\bm{q}_l}$  finally.

\par
\textbf{Warping and blending}.
The novel SAI at the target position $\bm{q}_l$ can be synthesized by warping  the input SAIs in $\mathcal{I}_s$ using the predicted disparity map $\bm{D}_{\bm{q}_l}$.
Specifically, the resulting image $\bm{I}_{\bm{q}_l\gets\bm{p}_k}$ by warping $\bm{I}_{\bm{p}_k}$ to the target position $\bm{q}_l$ can be expressed as  
\begin{equation}
    \bm{I}_{\bm{q}_l\gets\bm{p}_k} ( \bm{x} ) = \bm{I}_{\bm{p}_k}\left(\bm{x} + (\bm{q}_l-\bm{p}_k)\cdot \bm{D}_{\bm{q}_l}(\bm{x})\right).
\end{equation}

\par
Since the input SAIs contain valuable information of the scene from different viewpoints, they will contribute to the target SAI in different areas.
The warped images inevitably show artifacts around occlusion boundaries, 
and locations of the artifacts  vary among different source SAIs.
Direct combination of the images warped from different viewpoints by simple average or convolutional layers trained with the $\ell_1$/$\ell_2$ loss \cite{lfrec2016kalantari} will produce blurry effects, especially when the input SAIs have large disparities.
Therefore,  we propose a blending strategy to fuse the images warped from different input SAIs to generate the novel SAI 
by using adaptive dense confidence maps.
Specifically, the confidence maps are learned to indicate the pixel-wise accuracy of the images warped from different input SAIs.
Then it is expected that the more accurate regions can be selected to form the synthesized SAIs.
This strategy  properly handles the occlusion problem after warping and preserves clear textures in the synthesized novel SAI (see details in \ref{sec:ablation}).

\par
The $K$ confidence maps corresponding to the $K$ input SAIs, along with the disparity maps, are predicted by the final layer of the disparity estimation network.
It is feasible because the network has learnt the relationships between the input SAIs and implicitly modeled their relationships to the target SAI.
Then the blending can be formulated as:
\begin{equation}
    \widetilde{\bm{I}}_{\bm{q}_l} = \sum_{k=1}^{K} \bm{C}_k\odot \bm{I}_{\bm{q}_l\gets\bm{p}_k},
\end{equation}
where $\bm{C}_k$ is the confidence map for $k$-th input SAI, and $\odot$ is the element-wise multiplication operator.

\subsection{Efficient LF Refinement}
In the coarse SAI synthesis phase, novel SAIs are independently synthesized, and the LF parallax structure among them are not well taken into account,
resulting in possible photometric inconsistencies between SAIs in the intermediate LF image $\widehat{\mathcal{I}}$.
Therefore, an efficient refinement network is designed to further exploit the structure of $\widetilde{\mathcal{I}}$, which is expected to recover the photo-consistency and further improve the reconstruction quality of the densely-sampled LF. 
Since the goal is to correct possible flaws inconsistent across SAIs while preserve high-frequency textures, residual learning is used in this module.
In summary, we first exploit the LF parallax structure from $\widehat{\mathcal{I}}$ and then reconstruct residual maps for it, as formulated in Eq. $($\ref{eq:refine}$)$.

\par
\textbf{The LF parallax structure}.
To exploit the LF parallax structure within $\widetilde{\mathcal{I}}$, 4-D convolution is a straightforward choice. 
However, the computational cost required by 4-D convolution is very high. 
Instead, pseudo filters or separable filters, which reduce model complexity by approximating a high dimensional filter with a combination of filters of lower dimensions, have been applied to solve different computer vision problems,
such as image structure extraction \cite{sepfilter2013learning}, 3-D rendering \cite{sepfilter2015FSF} and video frame interpolation \cite{sepfilter2017video}. This has been recently adopted in \cite{sepfilter2016lf} for LF material classification and \cite{lfsr2019sas} for LF spatial super-resolution, which verifies that  pseudo 4-D filters can achieve comparable performance to 4-D filters.

\par
Therefore, we adopt the pseudo 4-D filter which approximates a single 4-D filtering step with two 2-D filters.
Specifically, the intermediate feature maps are reshaped between the stack of spatial feature maps $F_{spa} \in\mathbb{R}^{W\times  H\times f_c \times MN }$  and the stack of angular ones $F_{ang} \in\mathbb{R}^{M\times N\times f_c \times WH }$ so that the convolution is performed alternatively on the spatial and angular domains.
Such a design reduces the computation required by a 4-D convolution significantly, while it is still capable of extracting information from both spatial and angular information from the LF image effectively.

\par
\textbf{Residual reconstruction}.
After exploring the  relationship among angular dimension, the residual maps are reconstructed separately for each SAI in the intermediate LF image.
Several layers of 2-D spatial convolution are applied to learn a residual map from the extracted spatial-angular deep features for each SAI.  
Here each SAI is processed independently for two reasons.
First, we believe the previous spatial-angular convolutions are capable of exploiting the LF parallax structure.
Second and more importantly, in this way, we can build a fully-convolutional network on both spatial and angular dimension, such that flexible output angular resolution is achieved.
Finally, the reconstructed residual map is added to the previously synthesized intermediate LF image as the final reconstructed LF $\widehat{I}$.
    
    \begin{figure*}[!t]
    \centering
    \subfloat[$4\rightarrow7\times7$]{\includegraphics[width=0.33\linewidth]{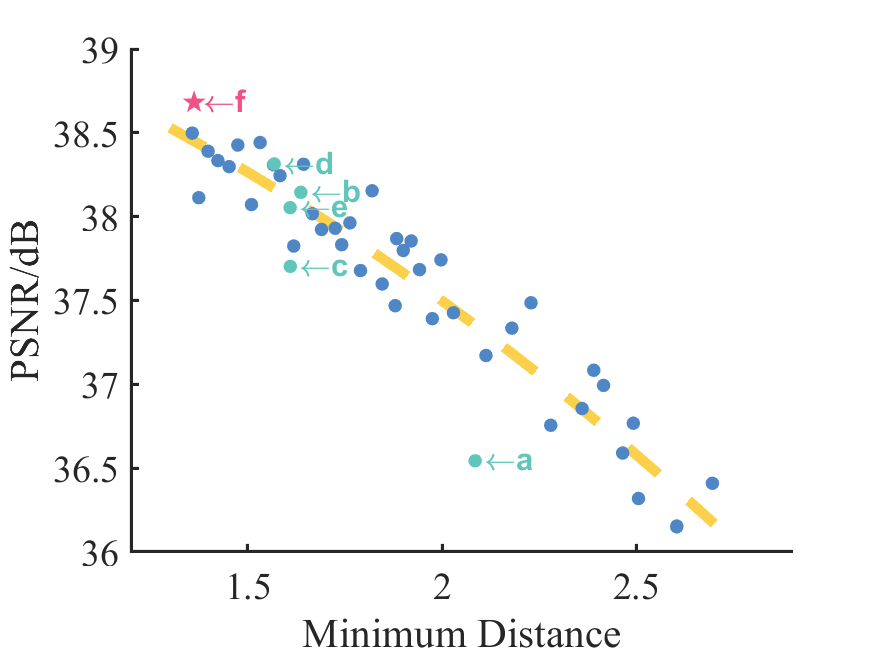}}
    \subfloat[$3\rightarrow7\times7$]{\includegraphics[width=0.33\linewidth]{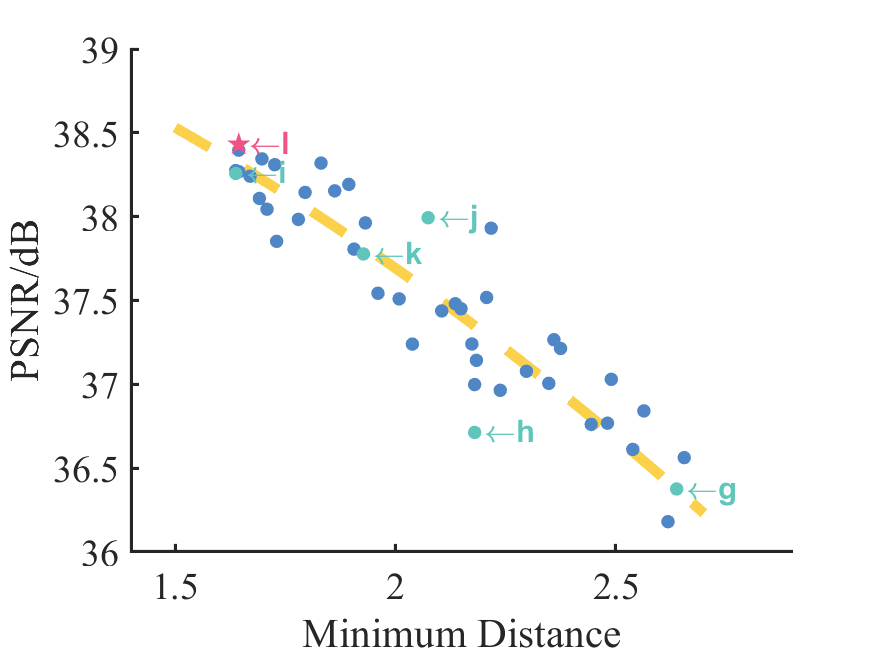}}
    \subfloat[$2\rightarrow7\times7$]{\includegraphics[width=0.33\linewidth]{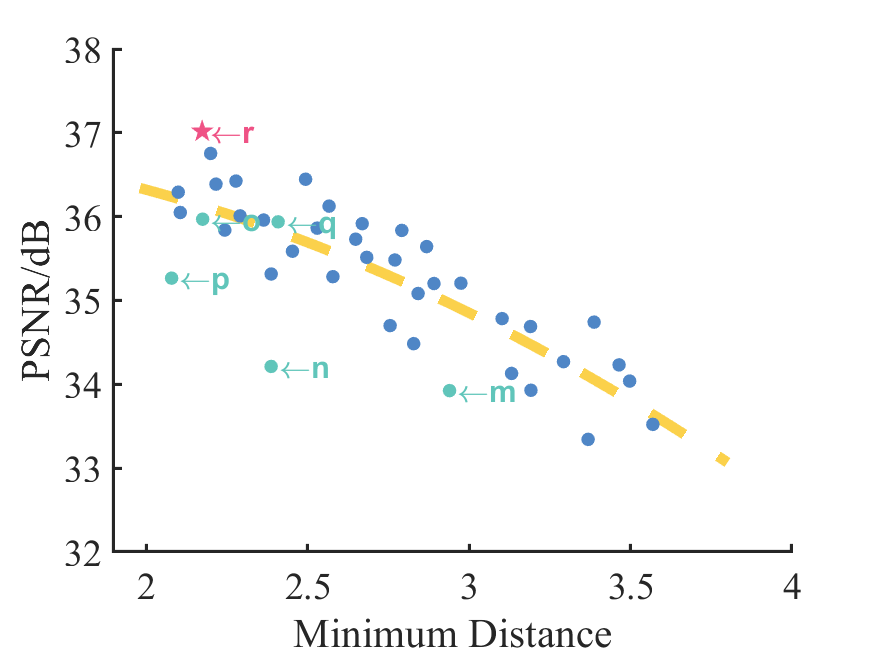}}
    \caption{Illustration of the relationship between the  minimum distance of the sampling patterns and the reconstruction quality tested on the \textit{HCI} dataset. 
    The blue dots denote the patterns generated randomly. The green dots and their annotations  correspond  to the patterns in Fig. \ref{fig_pattern}. The results of the optimized patterns by our method are highlighted as red stars.}
    \label{fig_plot_dist}
    \end{figure*}
    
    \begin{figure}[!t]
    \centering
    \subfloat[]{\includegraphics[width=0.15\linewidth]{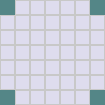}%
    \label{subfig_4a}}
    \hfil
    \subfloat[]{\includegraphics[width=0.15\linewidth]{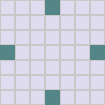}%
    \label{subfig_4b}}
    \hfil
    \subfloat[]{\includegraphics[width=0.15\linewidth]{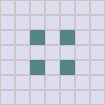}%
    \label{subfig_4c}}
    \hfil
    \subfloat[]{\includegraphics[width=0.15\linewidth]{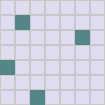}%
    \label{subfig_4d}}
    \hfil
    \subfloat[]{\includegraphics[width=0.15\linewidth]{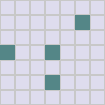}%
    \label{subfig_4e}}
    \hfil
    \subfloat[]{\includegraphics[width=0.15\linewidth]{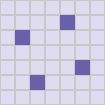}%
    \label{subfig_4f}}\\
    \subfloat[]{\includegraphics[width=0.15\linewidth]{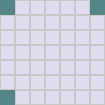}%
    \label{subfig_3g}}
    \hfil
    \subfloat[]{\includegraphics[width=0.15\linewidth]{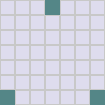}%
    \label{subfig_3h}}
    \hfil
    \subfloat[]{\includegraphics[width=0.15\linewidth]{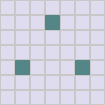}%
    \label{subfig_3i}}
    \hfil
    \subfloat[]{\includegraphics[width=0.15\linewidth]{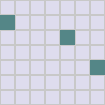}%
    \label{subfig_3j}}
    \hfil
    \subfloat[]{\includegraphics[width=0.15\linewidth]{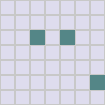}%
    \label{subfig_3k}}
    \hfil
    \subfloat[]{\includegraphics[width=0.15\linewidth]{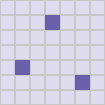}%
    \label{subfig_3l}}\\
    \subfloat[]{\includegraphics[width=0.15\linewidth]{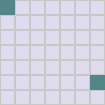}%
    \label{subfig_2m}}
    \hfil
    \subfloat[]{\includegraphics[width=0.15\linewidth]{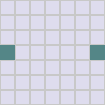}%
    \label{subfig_2n}}
    \hfil
    \subfloat[]{\includegraphics[width=0.15\linewidth]{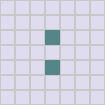}%
    \label{subfig_2o}}
    \hfil
    \subfloat[]{\includegraphics[width=0.15\linewidth]{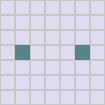}%
    \label{subfig_2p}}
    \hfil
    \subfloat[]{\includegraphics[width=0.15\linewidth]{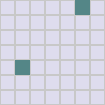}%
    \label{subfig_2q}}
    \hfil
    \subfloat[]{\includegraphics[width=0.15\linewidth]{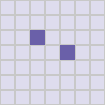}%
    \label{subfig_2r}}
    \caption{Illustration of different sampling patterns. 
    From top to bottom are sampling patterns with 4, 3 and 2 input SAIs, respectively.
    (f), (l) and (r) depict the optimized sampling patterns by our algorithm for the tasks $4\rightarrow7\times7$, $3\rightarrow7\times7$ and $2\rightarrow7\times7$, respectively.}
    \label{fig_pattern}
    \end{figure}

\subsection{The Loss Function}
All modules in our approach are differentiable, leading to an end-to-end trainable network.
The loss function for training the network consists of three parts.
The first part provides supervision for the intermediate LF by calculating  the absolute error between the intermediate LF images and ground-truth ones, i.e.,
\begin{equation}
    \ell_s =  \| \mathcal{I} - \widetilde{\mathcal{I}} \|_1.
\end{equation}
To promote smoothness of the predicted ray disparity, we penalize the $L_1$ norm of the  second-order gradients \cite{vijayanarasimhan2017sfm}, denoted as $\ell_{smooth}$:
\begin{equation}
\begin{aligned}
    \ell_{smooth} =  \sum_{l=1}^{MN-K} \| \nabla_{xx}\bm{D}_{\bm{q}_l} \|_1 
    + \| \nabla_{xy}\bm{D}_{\bm{q}_l} \|_1 \\
    + \| \nabla_{yx}\bm{D}_{\bm{q}_l} \|_1 + \| \nabla_{yy}\bm{D}_{\bm{q}_l} \|_1 ,
\end{aligned}
\end{equation}
where $\nabla_{xx}$, $\nabla_{xy}$, $\nabla_{yx}$ and $\nabla_{yy}$ are the second-order gradients for the spatial domain of the disparity map $\bm{D}_{\bm{q}_l}$.
Finally, the output reconstructed LF image is optimized by minimizing the absolute error as:
\begin{equation}
    \ell_r =  \| \mathcal{I} - \widehat{\mathcal{I}} \|_1.
\end{equation}
% \par
Thus, our final objective is written as
\begin{equation}
    \ell =  \lambda_1\ell_s + \lambda_2\ell_{smooth}+\lambda_3\ell_r,
\end{equation}
where $\lambda_1$, $\lambda_2$ and $\lambda_3$ are the weighting for the reconstruction accuracy and the disparity smoothness, which are empirically set to $1$, $0.001$ and $1$, respectively.

        \begin{figure*}[!t]
        \centering
        \includegraphics[width=\linewidth]{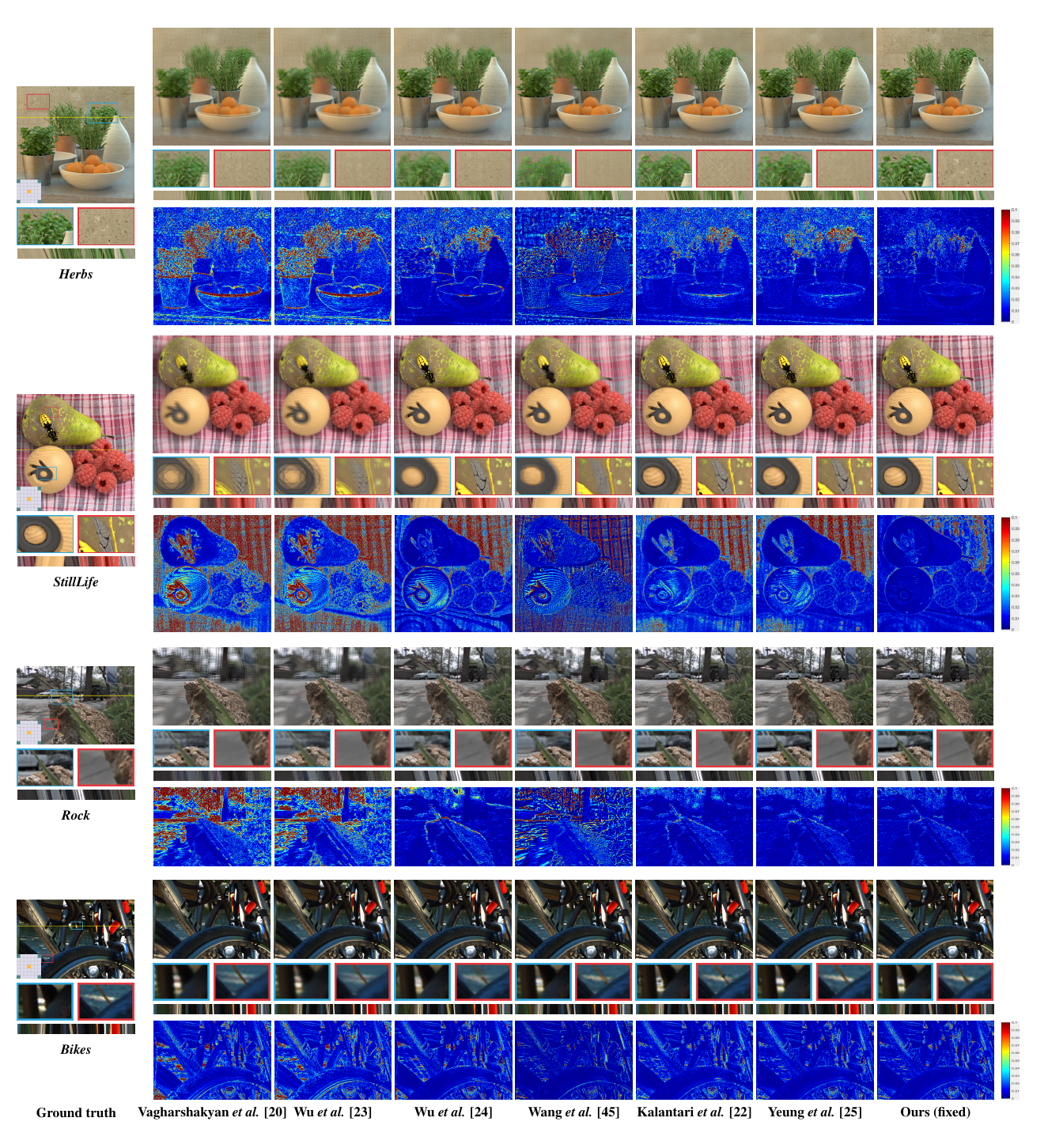}
        \caption{Visual comparisons of different methods on the synthesized center SAI for the task $2\times2\rightarrow7\times7$ (fixed models). Selected regions have been zoomed in for better comparison. It is recommended to view this figure by zooming in.}
        \label{fig_visual_fix}
        \end{figure*}

        \begin{table*}[htbp]%[htbp]
        \centering
        \caption{Comparison of attributes for densely-sampled LF reconstruction algorithms,
        % Qualitative comparisons of algorithms for densely-sampled LF reconstruction, 
        where \textit{flexible input} means whether the method is feasible for an arbitrary sampling pattern, and \textit{flexible output} means whether the method can produce densely-sampled  LFs with flexible angular resolution.}
        \resizebox{0.8\textwidth}{!}{
        \begin{tabular}{c|c c c c}
        \toprule[1.5pt]
        Algorithms & learning-based  & geometry-based & flexible input & flexible output\\
        \midrule[1pt]
        Vagharshakyan \textit{et al.} \cite{lfrec2018shearlet}  & - & - & - & \checkmark      \\
        Wu \textit{et al.} \cite{lfrec2018wu:blur}   &\checkmark  & - & - & \checkmark  \\
        Wu \textit{et al.} \cite{lfrec2019wu:shear}    & \checkmark  & \checkmark & - & \checkmark   \\
        Wang \textit{et al.} \cite{lfrec2018p4d} & \checkmark & - & - & - \\
        Kalantari \textit{et al.} \cite{lfrec2016kalantari}    & \checkmark & \checkmark  & \checkmark & \checkmark \\
        Yeung \textit{et al.} \cite{lfrec2018Yeung}  & \checkmark   & - & - &-  \\
        Ours   & \checkmark    & \checkmark & \checkmark & \checkmark \\
        \bottomrule[1.5pt]
        \end{tabular}
        }
        \label{tab:algorithms}
        \end{table*}

        \begin{table*}[!t]
        \renewcommand{\arraystretch}{1.3}
        \caption{Quantitative comparisons (PSNR/SSIM) of the proposed approach with the state-of-the-art ones under task $2\times2\rightarrow7\times7$. The input sparsely-sampled  LFs are sampled at the four corners during both training and test.
        }
        \label{table_fix}
        \centering
        \begin{tabular}{c|c|c c c c c |c c }
        \toprule[1.5pt]
        \multirow{2}{*}{Test set}  & \multirow{2}{*}{Disparity} & Vagharshakyan \textit{et al.}  & Wu \textit{et al.}  & Wu \textit{et al.}  & Wang \textit{et al.}  & Kalantari \textit{et al.} & Yeung \textit{et al.}  & \multirow{2}{*}{Ours (fixed)} \\
        ~ & ~ &  \cite{lfrec2018shearlet} & \cite{lfrec2018wu:blur} & \cite{lfrec2019wu:shear} & \cite{lfrec2018p4d} & \cite{lfrec2016kalantari}  & \cite{lfrec2018Yeung} & ~\\
        \midrule[1pt]
        \textit{HCI} & [-24, 24] & 26.98/0.734 & 26.64/0.744 & 31.84/0.898 & 29.61/0.819 & \underline{32.85/0.909} & 32.30/0.900  & \textbf{37.14}/\textbf{0.966}\\
        \textit{HCI old} & [-18, 18] & 32.47/0.853 & 31.43/0.850 & 37.61/0.942 & 35.73/0.898 &  38.58/0.944  & \underline{39.69/0.941} & \textbf{41.80}/\textbf{0.974} \\
        \midrule[1pt]
        \textit{30scenes} & [-6, 6] & 34.17/0.907 & 33.66/0.918 & 39.17/0.975 & 38.22/0.970 & 41.40/0.982 & \textbf{42.77}/\textbf{0.986} & \underline{42.75/0.986} \\
        \textit{Occlusions} & [-6, 6] & 32.64/0.923 & 32.72/0.924& 34.41/0.955 & 35.42/0.962 & 37.25/0.972 & \textbf{38.88}/\textbf{0.980} & \underline{38.51/0.979} \\
        \textit{Reflective} & [-6, 6] & 35.34/0.935 & 34.76/0.930 & 36.38/0.944 & 35.96/0.942 & 38.09/0.953 & \underline{38.33/0.960} & \textbf{38.35}/\textbf{0.957} \\
        \bottomrule[1.5pt]
        \end{tabular}
        \end{table*}

\subsection{Optimized Sampling Pattern}
Optimizing the sampling pattern for densely-sampled LF reconstruction is a valuable topic, which could further exploit the full potential of the reconstruction algorithm, and
improve the reconstruction quality using as few hardware resources as possible.
Additionally, optimizing the sampling pattern may be beneficial to its application in LF compression (see more details in Sec. \ref{sec:conclusion}).
In this section,
we first investigate how the sampling pattern affects the reconstruction qualitatively and experimentally,
then we propose a simple yet effective method for optimizing the sampling pattern tailored to our reconstruction model.

\par
Intuitively, the reconstruction quality is influenced by how thoroughly the scene has been recorded by the sparsely-sampled input. Since most foreground objects can be completely captured from different viewpoints, the occluded regions are the critical challenge. There are several factors that affect the amount of information that could be captured with LF over the occluded areas. One of the factors is the overall distance between the novel SAIs and the sampled SAIs. That is, SAIs nearby can provide more references for novel SAI reconstruction compared to those far away. Additionally, sampling patterns with SAIs distributed at more diverse locations along the horizontal and vertical directions are better than their counterparts with less variation, as the former sees more occluded regions.
Finally, this issue should be related to the scene content. Factors such as the geometry complexity between objects can play an important role.

\par
We experimentally investigated the effect of the sampling pattern on reconstruction quality. First, we define a metric, namely
minimum distance, which is the average of the angular Euclidean distances of all novel SAIs to their nearest input SAI in the 2-D sampling grid. 
We then conducted the following experiments, in which
we randomly selected some sampling patterns for $4\rightarrow7\times7$, $3\rightarrow7\times7$ and $2\rightarrow7\times7$ dense reconstruction, respectively, then fitted the  relationships between their minimum distance against their reconstruction quality with a second degree polynomial.
Fig. \ref{fig_plot_dist} illustrates the results, where we can see
that with the increase of the minimum distance of the sampling pattern, the corresponding reconstruction quality decreases in general.
Moreover, the corresponding sampling patterns of the green dots are illustrated in Fig. \ref{fig_pattern}. It can be seen that patterns with smaller variation along horizontal or vertical directions always stay below the fitted curve (e.g., with close values of the minimum distance, the sampling pattern $4(b)$ performs better than $4(c)$, and similar scenarios can be found between $3(l)$ and $3(i)$, and $2(q)$ and $2(n)$), which indicates that the divergence is indeed a factor influencing the reconstruction quality.

\par
Based on the above observations, 
we propose a simple yet effective strategy for optimizing the sampling pattern, which is formulated as:
\begin{equation}
\label{eq:pattern}
\begin{aligned}
     \mathop{\arg\min}_{\mathcal{P},\bm{O}} ~~ & \sum_{l=1}^{MN-K} \sum_{k=1}^{K}  o_{l,k} \parallel \bm{q}_l - \bm{p}_k \parallel_2^2,\\
     s.t.  ~ ~ & \bm{q}_l\in\mathcal{Q}, \, \bm{p}_k\in\mathcal{P},\\
      & \forall l,k, o_{l,k}\in[0,1],\\
      & \forall l, \sum_{k=1}^{K} o_{l,k}=1,
\end{aligned}
\end{equation}
where $o_{l,k}$ is the $(l,k)$-th entry of the indicator matrix $\bm{O}\in\mathbb{R}^{(MN-K)\times K}$, which indicates whether the $k$-th sampled SAI is the nearest one in all samples to the $l$-th novel SAI.
We first find a solution of the optimization problem in  Eq. $($\ref{eq:pattern}$)$ using the deterministic annealing based method \cite{hou2015human,lloyd1982kmeans}.
As the solution varies with initialization, we select the one producing the minimum objective value after repeating the algorithm with random initialization for 5 times.
In addition, as the resulting optimized positions may not be located on the grid, we consider the divergence along both horizontal and vertical directions to round the solutions.
In this way, we can obtain the optimized sampling patterns as depicted in Fig. \ref{fig_pattern}$(f)$, \ref{fig_pattern}$(l)$ and \ref{fig_pattern}$(r)$.
As demonstrated in Fig. \ref{fig_plot_dist}, 
the corresponding quantitative reconstruction quality under the sampling patterns by our algorithm achieves the highest when compared with others, which indicates the effectiveness of our sampling pattern selection algorithm.  
Furthermore, we experimentally verified the effectiveness of the proposed strategy for optimizing the sampling pattern on LFs with different scene content, see section \ref{sec:ablation}.
Note that flexible and optimized sampling is not applicable to the micro-lens-based LF camera with a fixed optical sampling pattern.

        \begin{table*}[!t]
        \renewcommand{\arraystretch}{1.3}
        \caption{Quantitative comparisons of the proposed approach with Kalantari \textit{et al.} \cite{lfrec2016kalantari} on the reconstruction with arbitrary sampling patterns under task $4\rightarrow7\times7$. Sampling patterns (a), (c) and (f) (depicted in Fig. \ref{fig_pattern}) are used for comparison.}
        \label{table_arb_S4}
        \centering
        \begin{tabular}{c| c c | c c | c c}
        \toprule[1.5pt]
         & \multicolumn{2}{c|}{$4 (a) \rightarrow 7\times7$} & \multicolumn{2}{c|}{$4(c) \rightarrow 7\times7$} & \multicolumn{2}{c}{$4(f) \rightarrow 7\times7$} \\
        \midrule[1pt]
        Test set  & Kalantari \textit{et al.} \cite{lfrec2016kalantari}   & Ours  & Kalantari  \textit{et al.} \cite{lfrec2016kalantari}   & Ours   & Kalantari \textit{et al.} \cite{lfrec2016kalantari}   & Ours  \\
        \midrule[1pt]
        \textit{HCI} & 32.22/0.908 & \textbf{36.54/0.961} & 33.98/0.929  & \textbf{37.70/0.966} & 34.28/0.933 & \textbf{38.68/0.971}\\
        \textit{HCI old} & 37.47/0.941 & \textbf{41.13/0.976} & 38.43/0.951 & \textbf{41.80/0.979} & 38.87/0.954 & \textbf{43.06/0.984} \\
        \midrule[1pt] 
        \textit{30scenes} & 40.06/0.978 & \textbf{41.18/0.982} & 40.72/0.981  & \textbf{42.02/0.985} & 40.91/0.982 & \textbf{42.83/0.986} \\
        \textit{Occlusions} & 35.17/0.962 & \textbf{36.45/0.970} & 36.90/ 0.971  & \textbf{38.45/0.977} & 36.88/0.971 & \textbf{39.57/0.981} \\
        \textit{Reflective} & 36.38/0.941  & \textbf{37.05/0.946} & 38.60/0.957 & \textbf{39.41/0.960} & 38.64/0.956 & \textbf{40.15/0.961}\\
        \bottomrule[1.5pt]
        \end{tabular}
        \end{table*}
        
        \begin{table*}[!t]
        \renewcommand{\arraystretch}{1.3}
        \caption{Quantitative comparisons of the proposed approach with Kalantari \textit{et al.} \cite{lfrec2016kalantari} on the reconstruction with arbitrary sampling patterns under task $3\rightarrow7\times7$. Sampling patterns (g), (j) and (l) (depicted in Fig. \ref{fig_pattern}) are used for comparison.}
        \label{table_arb_S3}
        \centering
        \begin{tabular}{c| c c | c c | c c}
        \toprule[1.5pt]
         & \multicolumn{2}{c|}{$3(g) \rightarrow 7\times7$} & \multicolumn{2}{c|}{$3(j) \rightarrow 7\times7$} & \multicolumn{2}{c}{$3(l) \rightarrow 7\times7$} \\
        \midrule[1pt]
        Test set  & Kalantari \textit{et al.} \cite{lfrec2016kalantari}   & Ours  & Kalantari \textit{et al.} \cite{lfrec2016kalantari}   & Ours  & Kalantari \textit{et al.} \cite{lfrec2016kalantari}   & Ours  \\
        \midrule[1pt]
        \textit{HCI} & 31.02/0.883 & \textbf{36.38/0.960} & 33.23/0.918 & \textbf{37.99/0.967} & 33.49/0.922 & \textbf{38.43/0.970}\\
        \textit{HCI old} & 36.33/0.927 & \textbf{41.22/0.976} & 38.02/0.947 & \textbf{42.48/0.981} & 38.49/0.949 & \textbf{43.09/0.983} \\
        \midrule[1pt] 
        \textit{30scenes} & 38.95/0.973 & \textbf{40.65/0.981} & 40.56/0.980 & \textbf{41.86/0.984} & 40.86/0.981 & \textbf{42.57/0.986} \\
        \textit{Occlusions} & 34.05/0.951 & \textbf{35.80/0.967} & 36.14/0.967 & \textbf{38.000.976} & 36.63/0.970 & \textbf{39.12/0.980} \\
        \textit{Reflective} & 35.49/0.936 & \textbf{36.43/0.948} & 38.30/0.951  & \textbf{39.41/0.958} & 38.77/0.954 & \textbf{40.00/0.961}\\
        \bottomrule[1.5pt]
        \end{tabular}
        \end{table*}

        \begin{table*}[!t]
        \renewcommand{\arraystretch}{1.3}
        \caption{Quantitative comparisons of the proposed approach with Kalantari \textit{et al.} \cite{lfrec2016kalantari} on the reconstruction with arbitrary sampling patterns under task $2\rightarrow7\times7$. Sampling patterns (m), (p) and (r) (depicted in Fig. \ref{fig_pattern}) are used for comparison.}
        \label{table_arb_S2}
        \centering
        \begin{tabular}{c| c c | c c | c c}
        \toprule[1.5pt]
         & \multicolumn{2}{c|}{$2(m) \rightarrow 7\times7$} & \multicolumn{2}{c|}{$2(p) \rightarrow 7\times7$} & \multicolumn{2}{c}{$2(r) \rightarrow 7\times7$} \\
        \midrule[1pt]
        Test set  & Kalantari \textit{et al.} \cite{lfrec2016kalantari}   & Ours & Kalantari \textit{et al.} \cite{lfrec2016kalantari}   & Ours  & Kalantari \textit{et al.} \cite{lfrec2016kalantari}   & Ours  \\
        \midrule[1pt]
        \textit{HCI} & 30.69/0.877 & \textbf{33.93/0.946} & 31.65/0.897  & \textbf{35.27/0.957} & 32.50/0.906 & \textbf{37.02/0.963}\\
        \textit{HCI old} & 36.05/0.927 & \textbf{40.44/0.967} & 36.27/0.933 & \textbf{39.88/0.961} & 36.46/0.939 & \textbf{41.30/0.977} \\
        \midrule[1pt]
        \textit{30scenes} & 37.42/0.964 & \textbf{40.05/0.979} & 38.83/0.974  & \textbf{40.79/0.981} & 38.54/0.973 & \textbf{40.98/0.982} \\
        \textit{Occlusions} & 32.95/0.936 & \textbf{35.11/0.960} & 34.88/0.958 & \textbf{36.69/0.970} & 34.83/0.958 & \textbf{37.08/0.971} \\
        \textit{Reflective} & 34.88/0.929  & \textbf{36.53/0.944} & 36.15/0.945  & \textbf{38.35/0.956} & 36.82/0.950 & \textbf{38.45/0.956}\\
        \bottomrule[1.5pt]
        \end{tabular}
        \end{table*}

\section{Experimental Results }
\label{sec:experiment}

\subsection{Datasets and Implementation Details}
Both synthetic LF images from  the 4-D LF benchmarks \cite{lfdataset2016hci} \cite{lfdataset2013hciold}
and  real-world LF images captured with a Lytro Illum camera provided by Standford Lytro LF Archive \cite{lfdataset2016stanford} and Kalantari \textit{et al.} \cite{lfrec2016kalantari} were employed to train and test.
Specifically, 20 synthetic images and 100 real-world images were used for training, while
9 synthetic data, including 4 LF images from the \textit{HCI} \cite{lfdataset2016hci} dataset and 5 LF images from the \textit{HCI old} \cite{lfdataset2013hciold} dataset,
and  3 datasets with 70 real-world LF images captured with a Lytro Illum camera were used for test, 
namely \textit{30scenes} \cite{lfrec2016kalantari}, \textit{Occlusions} \cite{lfdataset2016stanford} and \textit{Reflective} \cite{lfdataset2016stanford}.
These datasets cover several important factors in evaluating the methods for LF reconstruction.
Specifically, the synthetic datasets contain high-resolution textures to measure the ability of maintaining high-frequency details. The real-world datasets can  evaluate the performance of different methods under natural illumination and practical camera distortion. Moreover, the \textit{HCI} dataset contains LF images with large disparities, which emphasizes the robustness on more sparse sampling. The \textit{Occlusions} and \textit{Reflective} datasets focus on  challenging scenes in which  the assumption of photo-consistency is not guaranteed.

\par
During training, patches of spatial size $64\times 64$ were randomly cropped, and the batch size was set to 1 due to the limitation of the computational memory.
Moreover, we adopted ADAM \cite{kingma2014adam} optimizer with $\beta_1=0.9$ and $\beta_2=0.999$. The learning rate was initialized as $1e-4$ and reduced by a half when the loss stops decreasing.
The spatial resolution of the model output was kept unchanged at $64\times 64$ with padding of zeros.
We implemented the model with PyTorch. The code will be publicly available.

        \begin{figure*}[!t]
        \centering
        \includegraphics[width=\linewidth]{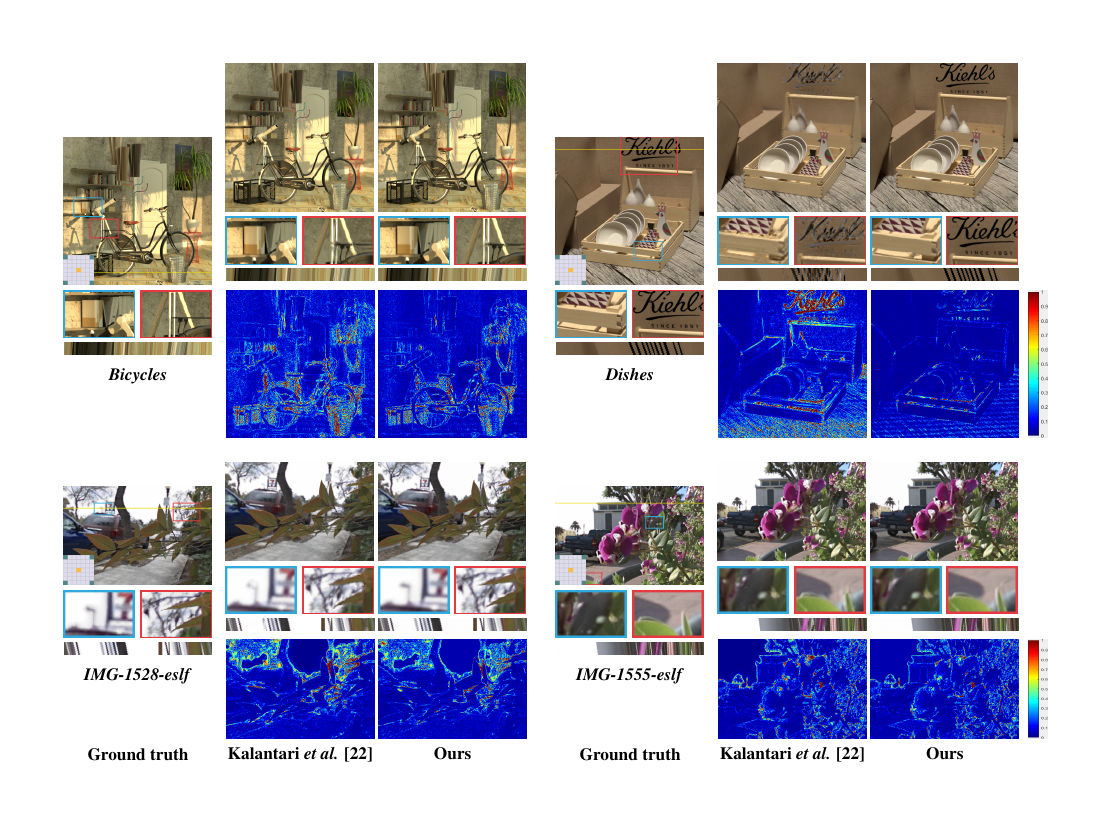}
        \caption{Visual comparisons of different methods on the synthesized center SAI for the task $4(a) \rightarrow 7\times7$ (flexible models). Selected regions have been zoomed in for better comparison. It is recommended to view this figure by zooming in.}
        \label{fig_visual_arb}
        \end{figure*}

\subsection{Comparison with State-of-the-Art Methods}

Besides our preliminary work Yeung \textit{et al.} \cite{lfrec2018Yeung}, we also 
compared with 5 state-of-the-art learning-based methods that were specifically designed for densely-sampled LF reconstruction, i.e., Vagharshakyan  \textit{et al.} \cite{lfrec2018shearlet}, Wu \textit{et al.} \cite{lfrec2018wu:blur}, Wu \textit{et al.} \cite{lfrec2019wu:shear}, Wang \textit{et al.} \cite{lfrec2018p4d}, and Kalantari \textit{et al.} \cite{lfrec2016kalantari} \footnote{Note that the methods with training code released, i.e., Wu \textit{et al.} \cite{lfrec2019wu:shear}, Wang \textit{et al.} \cite{lfrec2018p4d}, Kalantari \textit{et al.} \cite{lfrec2016kalantari}, and Yeung \textit{et al.} \cite{lfrec2018Yeung} were retrained with the same training data for fair comparisons. The retrained models achieve comparable performance to those provided by the authors. For method without training code released, i.e. Wu \textit{et al.} \cite{lfrec2018wu:blur}, we used the trained model provided by the authors.  }.
Table \ref{tab:algorithms} lists the feature comparisons of these algorithms in terms of whether they are learning-based, geometry-based, whether they are flexible with arbitrary input patterns, and whether they can produce the reconstruction with flexible angular resolution.
We conducted various experiments for comparisons, listed as follows:
\begin{itemize}
\item as 5 out of 6 methods under comparison, i.e. Vagharshakyan  \textit{et al.} \cite{lfrec2018shearlet}, Wu \textit{et al.} \cite{lfrec2018wu:blur}, Wu \textit{et al.} \cite{lfrec2019wu:shear}, Wang \textit{et al.} \cite{lfrec2018p4d}, and Yeung \textit{et al.} \cite{lfrec2018Yeung}, are unable to handle the input with flexible and irregular sampling patterns, we first designed the experiment $2\times2\rightarrow7\times7$, in which the same and fixed sampling pattern was used during both training and testing, such that all compared methods can be evaluated. 
We name our method \textit{Ours (fixed)} under such a training setting. See subsection 1);
\item as both Ours and Kalantari \textit{et al.} \cite{lfrec2016kalantari} can accept flexible and irregular sampling patterns, we designed the experiments $4\rightarrow7\times7$, $3\rightarrow7\times7$ and $2\rightarrow7\times7$, in which sparsely-sampled  LFs each containing $K$ SAIs with arbitrary positions and structures were fed into the network during training, and some of patterns illustrated in Fig. \ref{fig_pattern}  were used during testing. 
Here we considered three cases, i.e., $K=2,3,4$, respectively.
See subsection 2); and
\item 
we also evaluated the running time for different methods. See subsection 3).
\end{itemize}

\par
\textbf{$1)$ Comparison on the reconstruction with fixed input sampling patterns.}

\par 
This comparison was performed over the task $2\times2\rightarrow7\times7$, 
which attempts to reconstruct a densely-sampled  LF with $7\times7$ SAI from a sparsely-sampled  LF with $2\times2$ SAIs distributed regularly. Here the SAIs of a sparsely-sampled  LF are located at the four corners of the densely-sampled  LF to be reconstructed, as shown in Fig. \ref{subfig_4a}.
We used the average value of PSNR and SSIM over all synthetic novel SAIs to quantitatively measure the quality of reconstructed densely-sampled  LFs,
and the corresponding results are listed in Table \ref{table_fix}.
It can be observed that:
\begin{itemize}
\item  the performance of all methods decreases when the disparity between input SAIs increases;
\item  EPI-based methods, including Vagharshakyan \textit{et al.} \cite{lfrec2018shearlet}, Wu \textit{et al.} \cite{lfrec2018wu:blur}, Wu \textit{et al.} \cite{lfrec2019wu:shear}, and Wang \textit{et al.} \cite{lfrec2018p4d}, are inferior to others.
The possible reason is that only 2 rows or columns of pixels are available during the reconstruction of each EPI, making it difficult to recover the intermediate linear structures without modeling the 2-D spatial structure, especially when the scenes are complicated. Among them, Wu \textit{et al.} \cite{lfrec2019wu:shear} performs relatively better, as depth information is utilized as guidance;
\item Kalantari \textit{et al.} \cite{lfrec2016kalantari} achieves good results on real-world datasets, which indicates the effectiveness of geometry-based warping. However, it fails on the \textit{HCI} dataset with larger disparities. 
The reason is that Kalantari \textit{et al.} \cite{lfrec2016kalantari} 
uses hand-crafted features to estimate the disparity and simple convolutional layers to combine the warped images, which makes it difficult to build long distance connection between SAIs with large disparities;
\item Yeung \textit{et al.} \cite{lfrec2018Yeung} achieves the best results on the real-world datasets, indicating that the pseudo 4-D filters effectively explore the spatial and angular  relationships between input SAIs.
However, this method also does not work well on the \textit{HCI} dataset, because it entirely relies on deep regression for novel view synthesis, which indicates the importance of explicit geometric modeling for the reconstruction based on sparse sampling; and
\item our approach achieves the highest PSNR/SSIM for the \textit{HCI} and \textit{HCI old} datasets, and comparable performance with Yeung \textit{et al.} \cite{lfrec2018Yeung} at \textit{30scenes}, \textit{Occlusions} and \textit{Reflective} datasets, showing the advantages of the proposed framework.
\end{itemize}

\par 
We also visually compared the reconstruction results of different algorithms, as shown in Fig. \ref{fig_visual_fix}. It can be observed that Wu \textit{et al.} \cite{lfrec2018wu:blur}, Wu \textit{et al.} \cite{lfrec2019wu:shear} and Wang \textit{et al.} \cite{lfrec2018p4d} fail to recover delicate structures, such as the leaves and the textures on the wall, while
Kalantari \textit{et al.} \cite{lfrec2016kalantari} and Yeung \textit{et al.} \cite{lfrec2018Yeung} struggle with large disparities. In contrary, our approach produces accurate estimations, which are closer to the ground-truth ones.

Moreover, the most valuable information of LF images is the LF parallax structure, which implicitly represents the scene geometry.
In Figs. \ref{fig_visual_fix}, and \ref{fig_visual_arb}, we visualized the EPIs of reconstructed LF images to compare the ability of different reconstruction methods on the preservation of the LF parallax structure.  It can be seen that the EPIs of \textit{Ours (fixed)} preserve clearer linear structures, which are closer to the ground truth ones.
Moreover, the advantage of our method on preserving the LF parallax structure is also quantitatively and qualitatively demonstrated in Sec. \ref{sec:depth_enhanc}, where the depth maps estimated from reconstructed LFs by different methods are compared \cite{LFmetric2019depth}.
Finally, we provided supplementary videos for
animations of reconstructed LFs to visually evaluate the view consistency (see https://github.com/jingjin25/LFASR-FS-GAF).

        \begin{table*}[!t]
        \renewcommand{\arraystretch}{1.3}
        \caption{Comparisons of the running time (in seconds) of different methods for reconstructing a densely-sampled  LF.}
        \label{table_time}
        \centering
        \begin{tabular}{c|c c c c c c c}
        \toprule[1.5pt]
        \multirow{2}{*}{Algorithms} & Vagharshakyan  \textit{et al.}  & Wu \textit{et al.}  & Wu \textit{et al.}  & Wang \textit{et al.} & Kalantari \textit{et al.}  & Yeung \textit{et al.}  & Ours (fixed) \\
        ~ & \cite{lfrec2018shearlet} & \cite{lfrec2018wu:blur} &  \cite{lfrec2019wu:shear} & \cite{lfrec2018p4d} & \cite{lfrec2016kalantari} & \cite{lfrec2018Yeung} & ~\\
        \midrule[1pt]
        \textit{HCI} $2\times2\rightarrow7\times7$ & 924.52 & 257.70 & 101.70 & 1.07 & 168.86 & 0.85  & 40.21 \\
        \bottomrule[1.5pt]
        \end{tabular}
        \end{table*}

\par
\textbf{$2)$ Comparison on the reconstruction with flexible input sampling patterns.}

\par
We performed comparisons over random input positions with Kalantari \textit{et al.} \cite{lfrec2016kalantari} and our approach.
During training, the input SAIs were selected at random positions, and the input patterns illustrated in Fig. \ref{fig_pattern} were used for testing.
We report the quantitative results of task $4\rightarrow7\times7$, $3\rightarrow7\times7$ and $2\rightarrow7\times7$ in Table \ref{table_arb_S4}, \ref{table_arb_S3} and \ref{table_arb_S2}, respectively. It can observed that our method improves the PSNR by around 4 dB on synthetic datasets and around 0.4-1 dB on real-world datasets.

\par
To visually compare the outputs from  Kalantari \textit{et al.} \cite{lfrec2016kalantari} with our method,
we calculated the error maps of the reconstructed center SAI under task $4(a) \rightarrow 7\times7$ in Fig. \ref{fig_visual_arb}.
The results further demonstrate the advantages of our proposed approach. As shown in the results of synthetic data in Fig. \ref{fig_visual_arb} (see the first row), basic textures are severely blurred or distorted in the reconstructed SAI of Kalantari \textit{et al.} \cite{lfrec2016kalantari} when the inputs have large disparities, while our method can reconstruct most of the high-frequency details. 
For real-world LF reconstruction in Fig. \ref{fig_visual_arb} (see the second row), Kalantari \textit{et al.} \cite{lfrec2016kalantari} produces artifacts near the boundaries of the foreground objects, while fine edges and small objects are well preserved in the results by our method.

    \begin{figure}[!t]
    \centering
    \includegraphics[width=\linewidth]{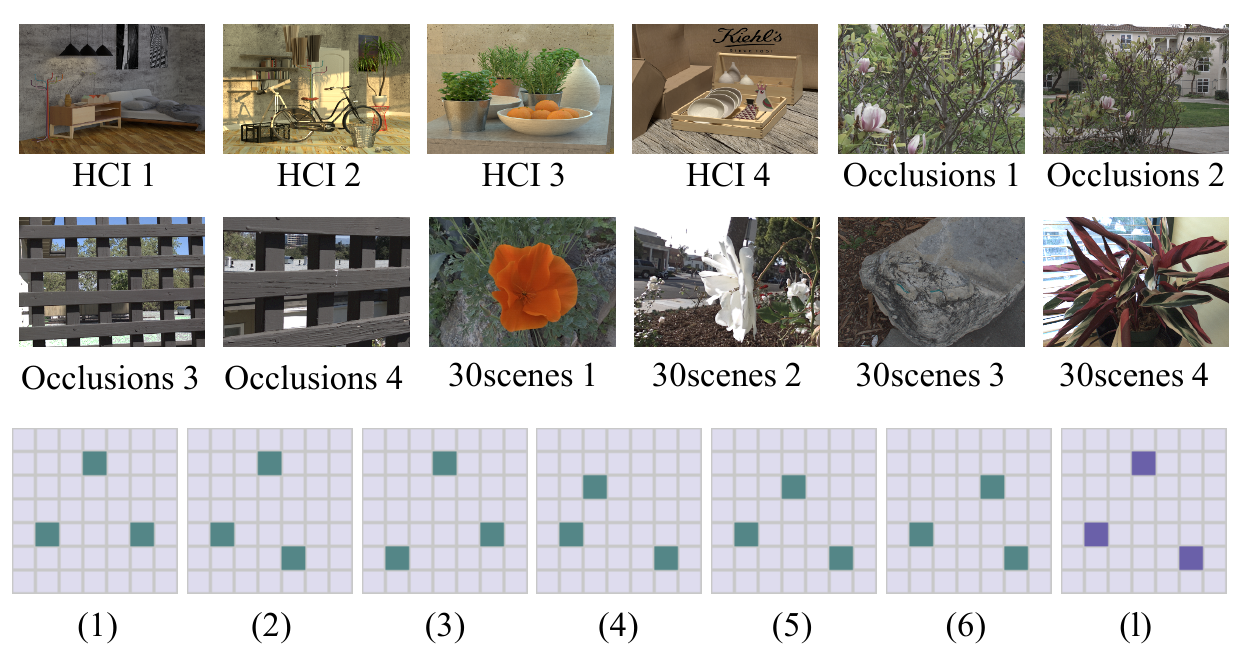}
    \caption{The images and sampling patterns used to investigate the effectiveness of the optimized sampling patterns on LFs with different scene content. 12 different scenes are manually selected. The optimized sampling pattern (l) obtained by our method is compared with 6 neighboring patterns (1)-(6).
    }
    \label{fig_pattern_content_img}
    \end{figure}
    
    \begin{figure}[!t]
    \centering
    \includegraphics[width=0.32\linewidth]{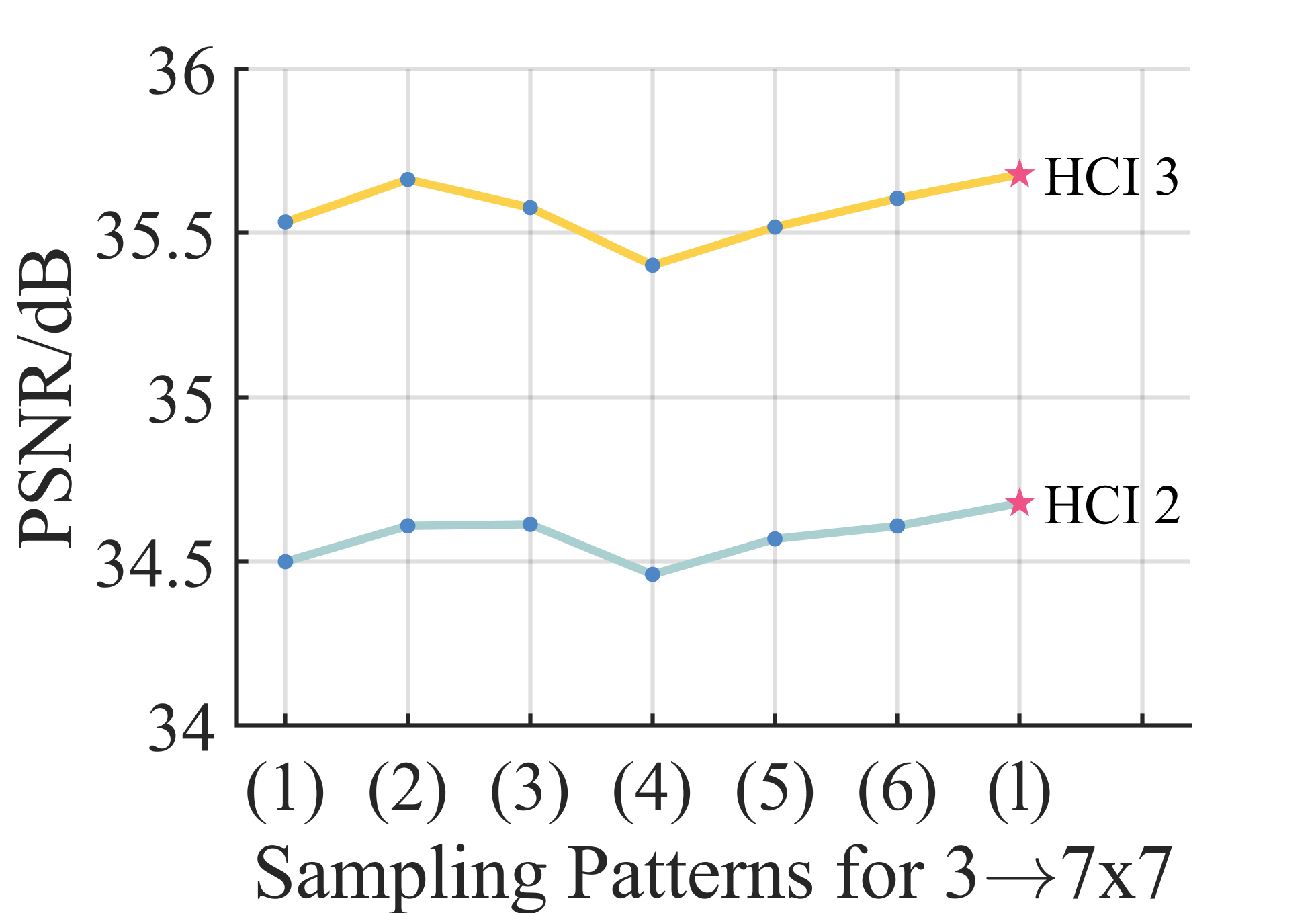}
    \includegraphics[width=0.32\linewidth]{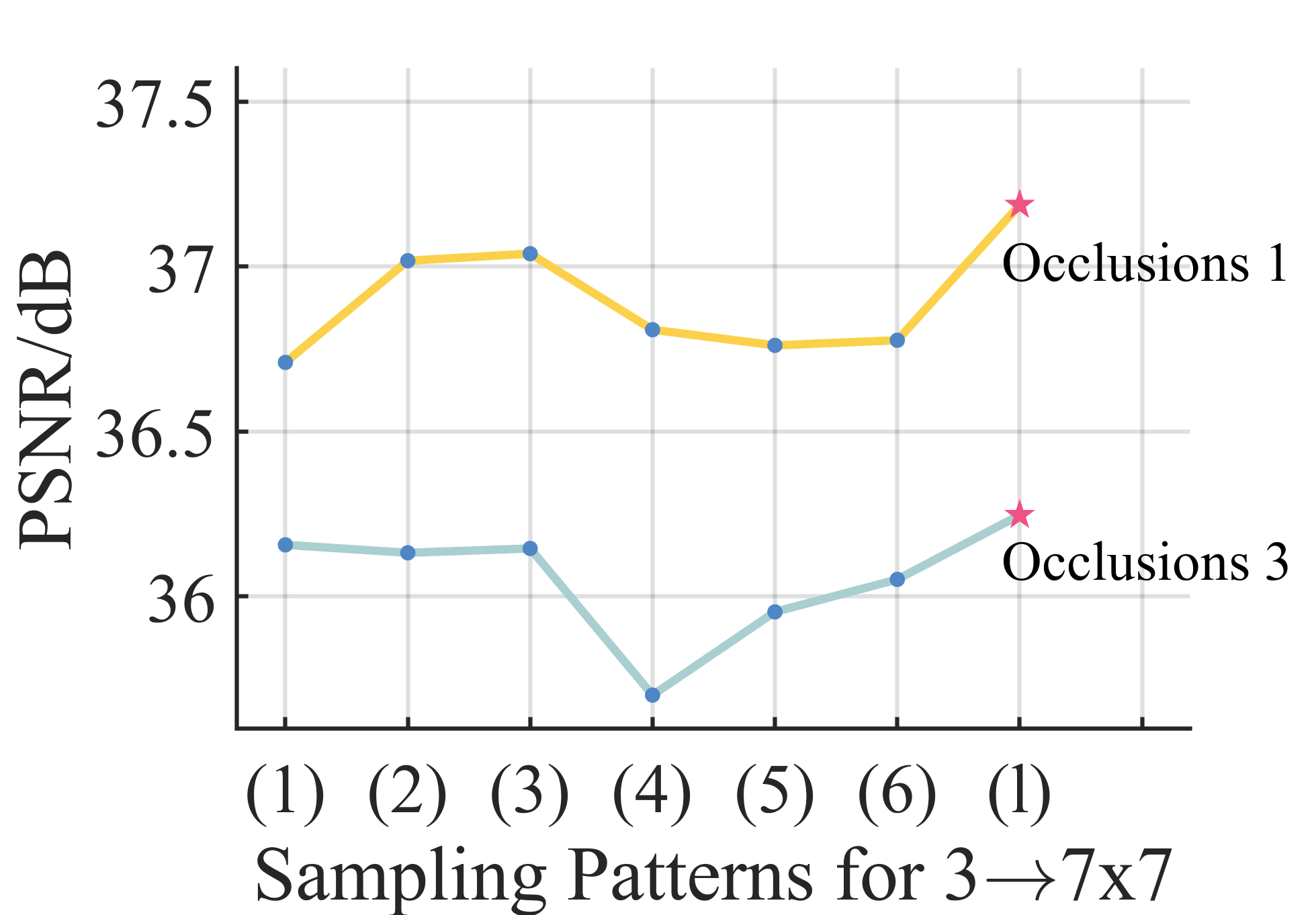}
    \includegraphics[width=0.32\linewidth]{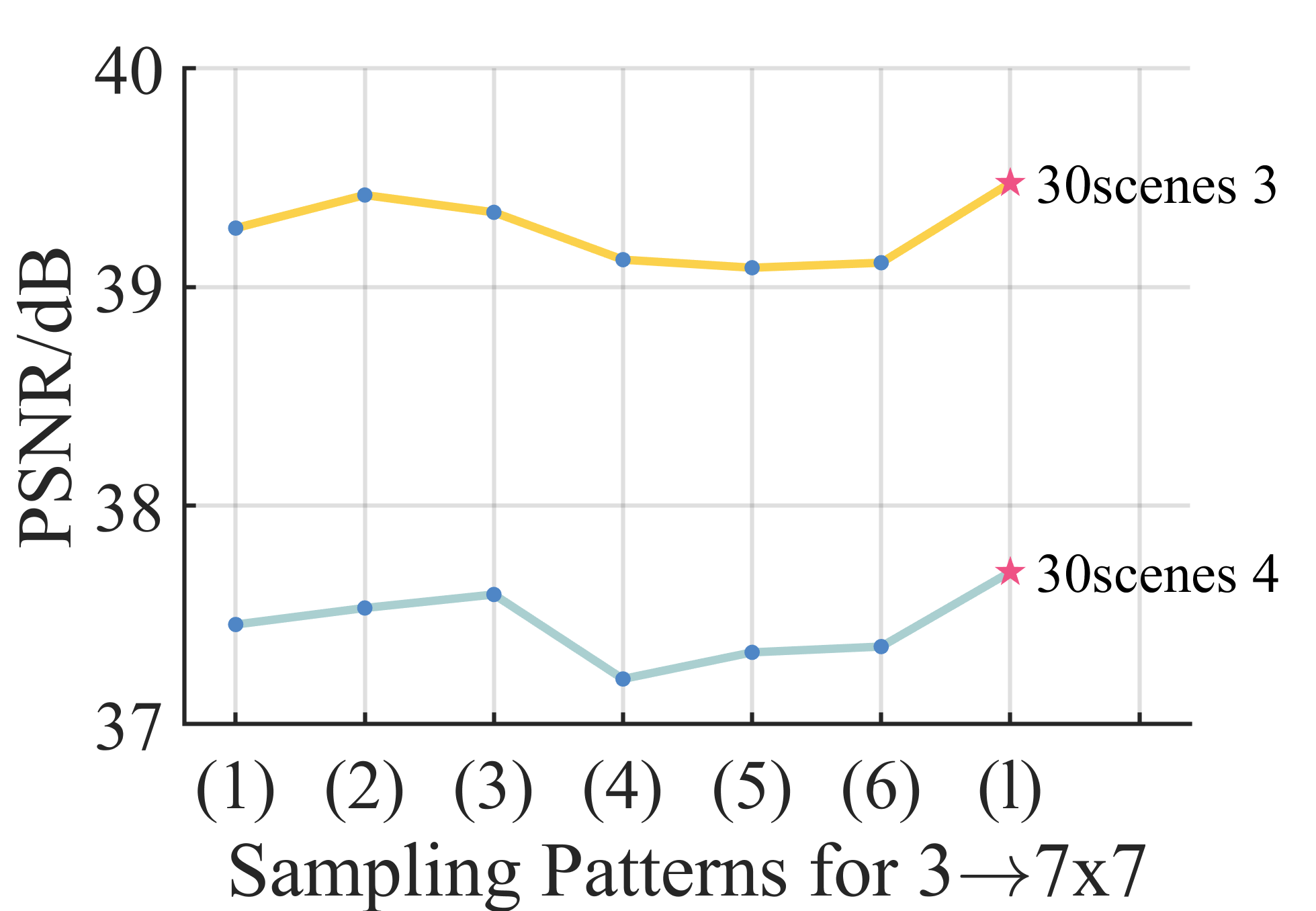}\\
    \includegraphics[width=0.32\linewidth]{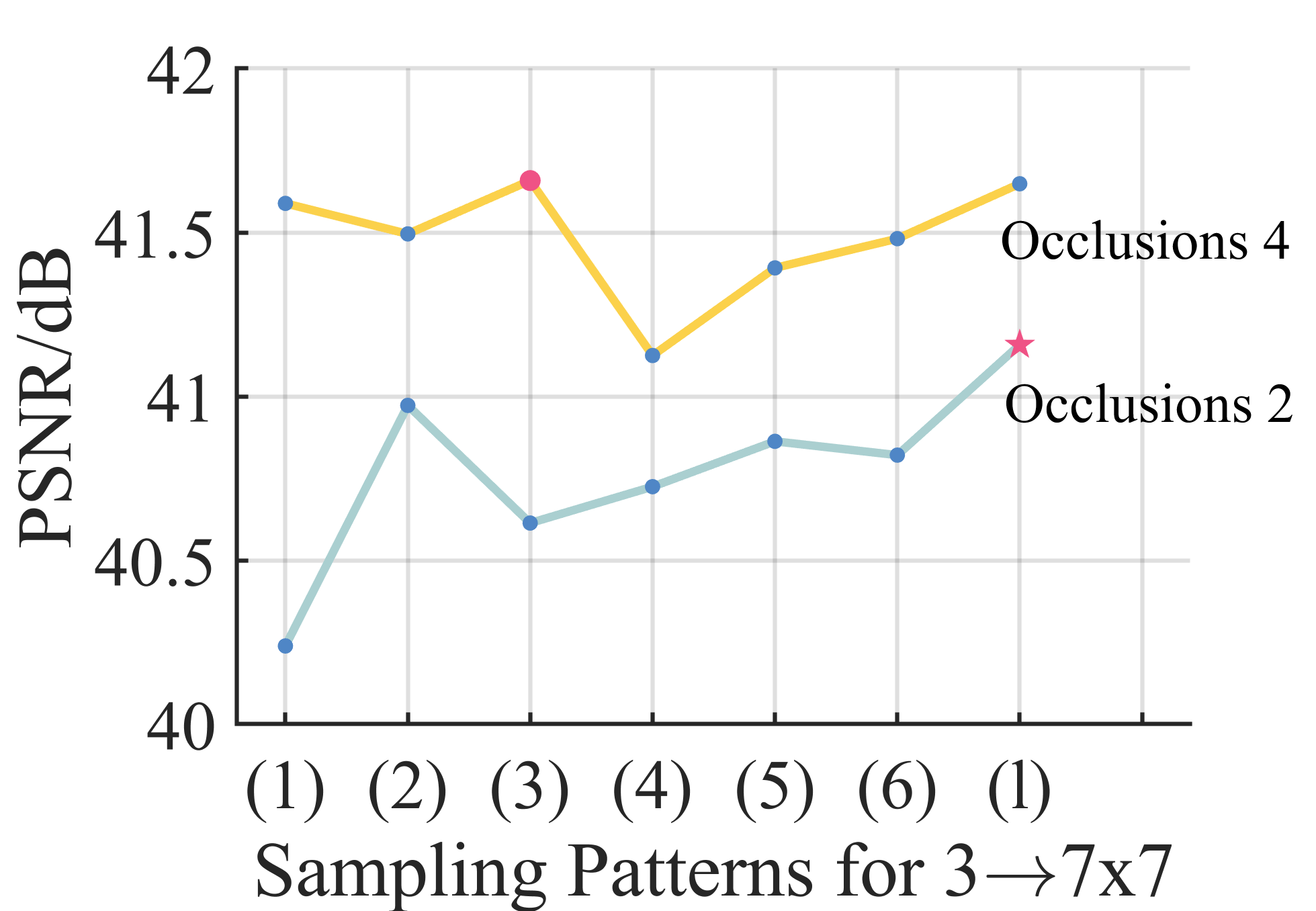}
    \includegraphics[width=0.32\linewidth]{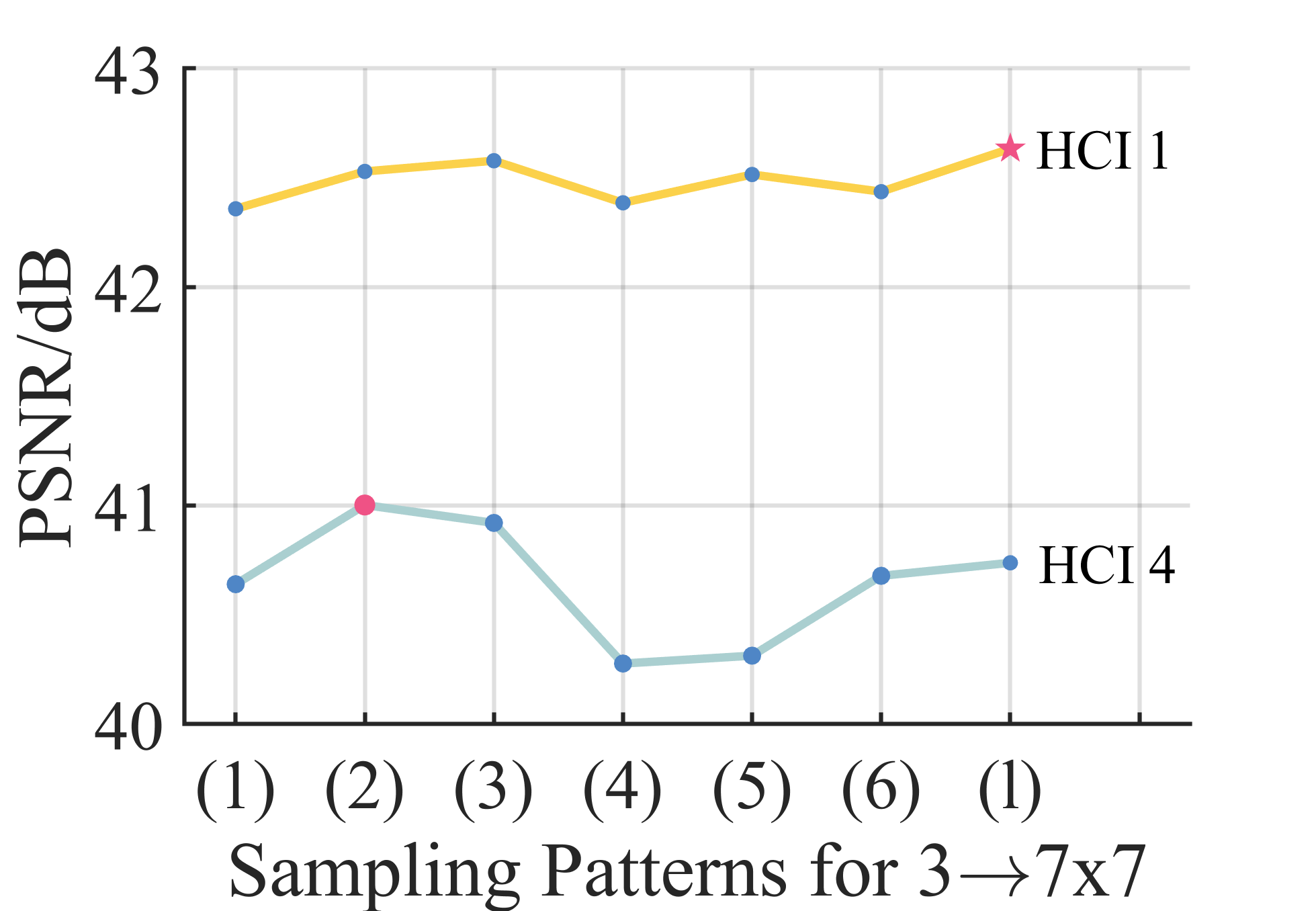}
    \includegraphics[width=0.32\linewidth]{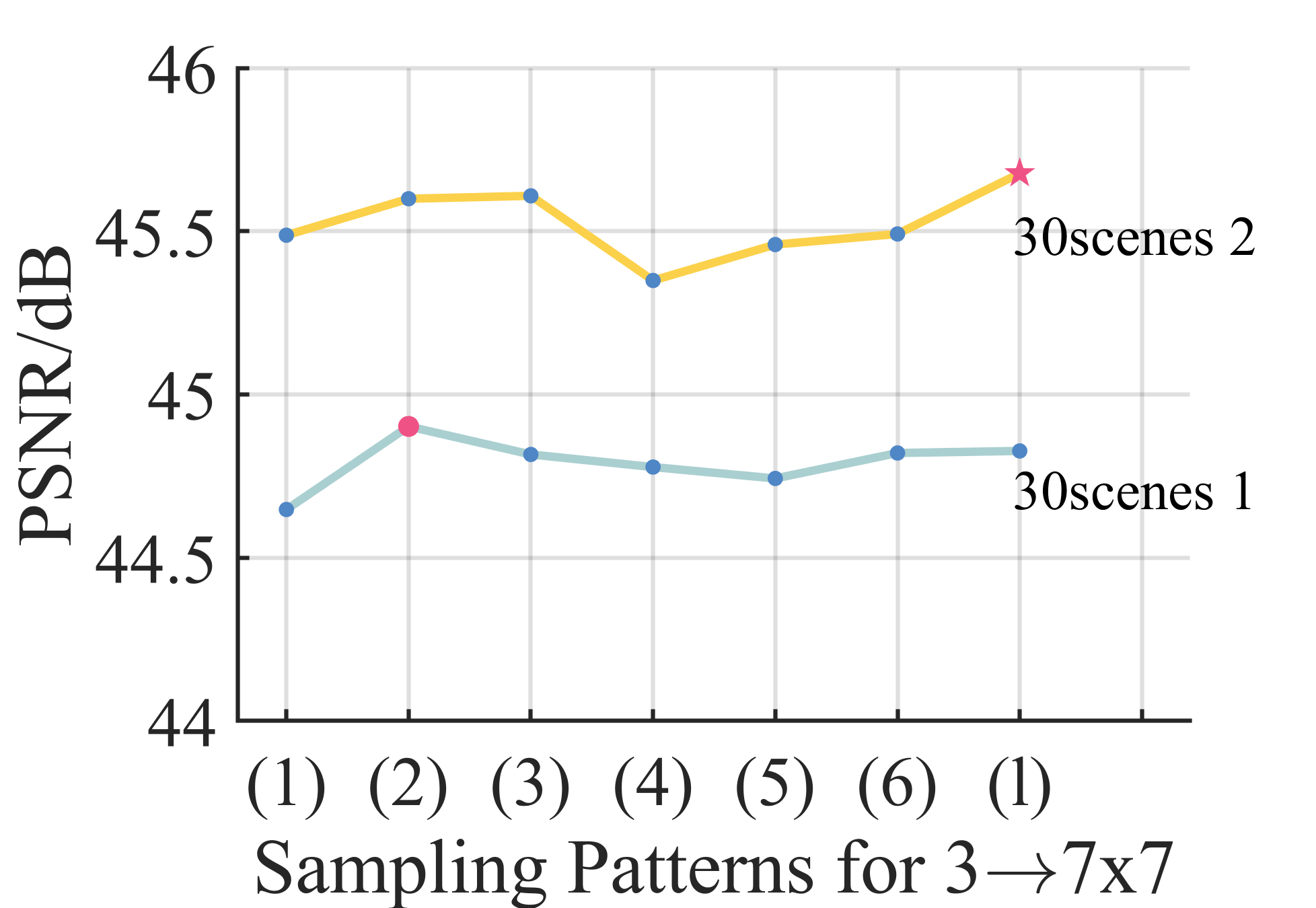}
    \caption{Illustration of the effectiveness of the optimized sampling patterns on LFs with different scene content. 
    The selected LF scenes and sampling patterns are illustrated in Fig. \ref{fig_pattern_content_img}.
    The red pentagrams mark the highest PSNR achieved with the optimized sampling pattern by our method, and red dots mark the highest PSNR achieved with other patterns.
    }
    \label{fig_pattern_content_psnr}
    \end{figure}

    \begin{figure*}[!t]
    \centering
    \includegraphics[width=0.8\linewidth]{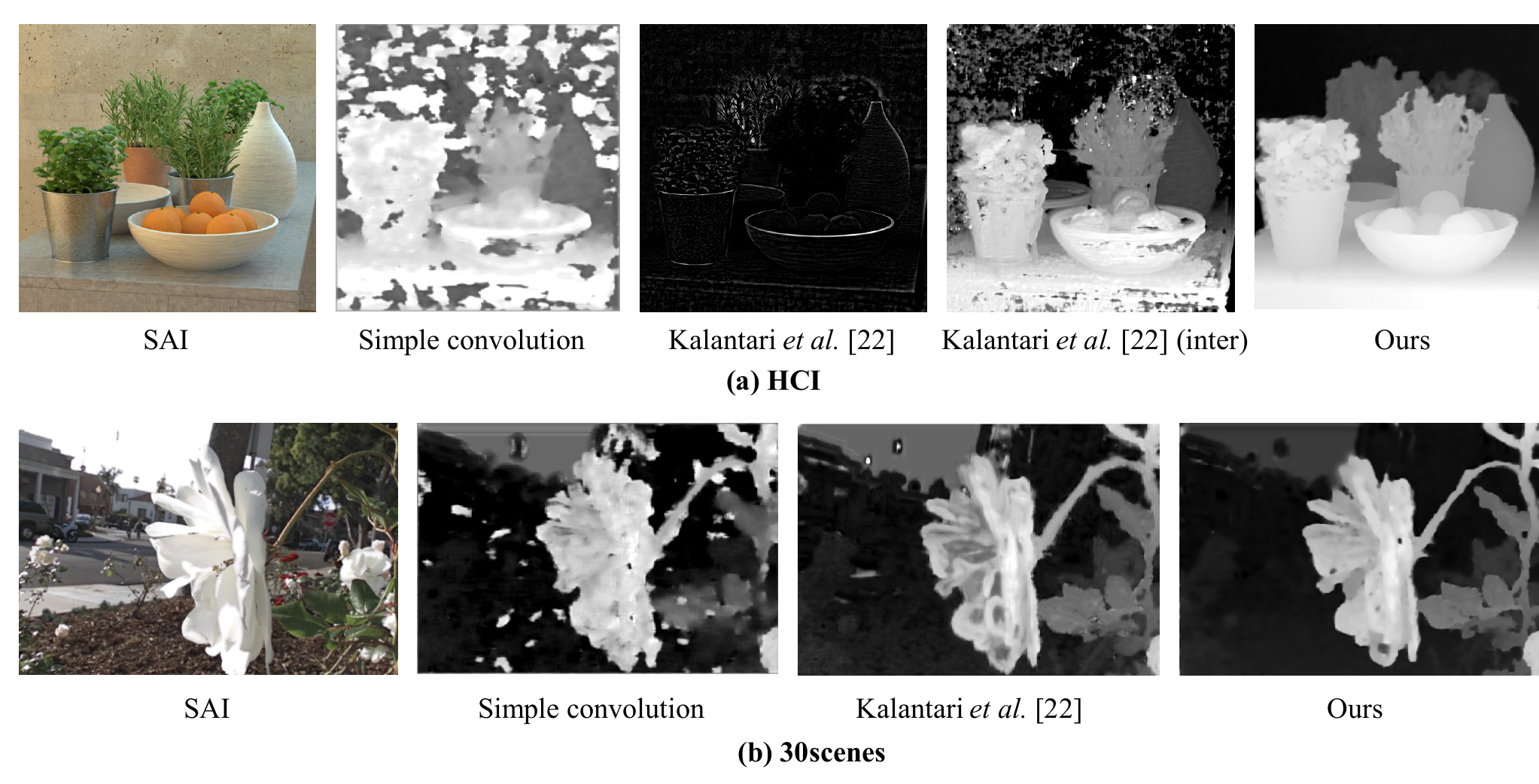}
    \caption{Visual comparisons of the intermediate by-product disparity maps estimated by directly applying convolutional layers to the input SAIs, Kalantari \textit{et al.} \cite{lfrec2016kalantari} and our network.
    Kalantari \textit{et al.} \cite{lfrec2016kalantari} (inter) denotes the modified network of  Kalantari \textit{et al.} \cite{lfrec2016kalantari} with an intermediate supervision for the warped images using ground-truth targets.
    }
    \label{fig_disp_inter}
    \end{figure*}

    \begin{figure*}[!t]
    \centering
    \includegraphics[width=0.8\linewidth]{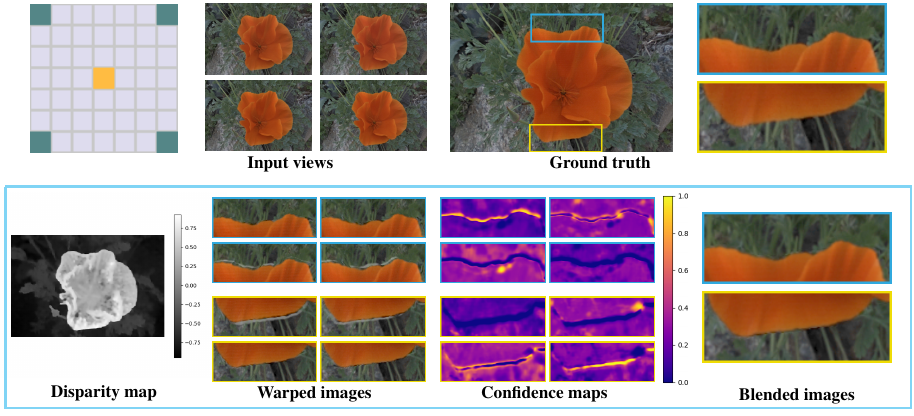}
    \caption{Demonstration of the effectiveness of our blending strategy.
    The estimated disparity map, the zoom-in of the images warped from the input SAIs, the learned confidence maps and the blended images are presented. }
    \label{fig_blend}
    \end{figure*}

\par
\textbf{$3)$ Comparison of the running time.}

\par
We compared the running time (in seconds) of different methods for reconstructing a densely-sampled  LF, and Table \ref{table_time} lists the results.
All methods were tested on a desktop with Intel CPU i7-8700 @ 3.70GHz, 32 GB RAM and NVIDIA GeForce RTX 2080 Ti.
From Table \ref{table_time}, it can be observed that our approach, taking about only 0.8 seconds to generate a novel SAI, is much faster than other methods except Wang \textit{et al.} \cite{lfrec2018p4d} and Yeung \textit{et al.} \cite{lfrec2018Yeung}.
Although Wang \textit{et al.} \cite{lfrec2018p4d} and Yeung \textit{et al.} \cite{lfrec2018Yeung} are the faster ones, our approach is superior in terms of reconstruction quality and angular flexibility.

    \begin{table}[!t]
    \renewcommand{\arraystretch}{1.3}
    \caption{Effectiveness verification of the refinement module in our approach. We compare the reconstruction quality of the LF images generated by our method without the refinement module and the LF images by our method with all modules under tasks $4\rightarrow7\times7$ and $3\rightarrow7\times7$ over \textit{HCI} and \textit{30scenes}.}
    \label{table_refine}
    \centering
    \resizebox{\linewidth}{!}{
    \begin{tabular}{c|c c | c c}
    \toprule[1.5pt]
    Test set  & without refinement & with refinement  & without refinement & with refinement  \\
    \midrule[1pt]
     & \multicolumn{2}{c|}{$4(a) \rightarrow 7\times7$} & \multicolumn{2}{c}{$4(f) \rightarrow 7\times7$} \\
    \textit{HCI} & 35.60/0.954 & \textbf{36.54/0.961} & 37.33/0.965 & \textbf{38.68/0.971} \\
    \textit{30scenens} & 40.12/0.979 & \textbf{41.18/0.982} & 41.57/0.983 & \textbf{42.83/0.986}\\
    \midrule[1pt]
     & \multicolumn{2}{c|}{$3(g) \rightarrow 7\times7$} & \multicolumn{2}{c}{$3(l) \rightarrow 7\times7$} \\
     \textit{HCI} & 35.39/0.953 & \textbf{36.38/0.960} & 37.15/0.963 & \textbf{38.43/0.970}\\
    \textit{30scenens} & 39.77/0.977 & \textbf{40.65/0.981} & 41.49/0.983 & \textbf{42.57/0.986}\\
    \bottomrule[1.5pt]
    \end{tabular}
    }
    \end{table}

\subsection{Ablation study}
\label{sec:ablation}
In this section, we experimentally validated the effectiveness of our view sampling optimization strategy on LFs with different scene content, as well as
the effectiveness of three components of our network, including the disparity estimation module, the blending strategy and the refinement module.

\par
\textbf{$1)$ The effectiveness of the optimization strategy for the sampling pattern on different scene content.}

\par
As there is no metric to quantify the scene content complexity, we manually select images covering different scenes and captured with different camera settings. As shown in Fig. \ref{fig_pattern_content_img}, the selected images vary in geometry complexity (e.g. \textit{30scenes 2} and \textit{30scenes 3}), object category (e.g. \textit{Occlusions 1} and \textit{Occlusions 3}), camera parameters (e.g. \textit{Occlusions 3} and \textit{Occlusions 4}), and data acquisition method  (e.g. \textit{HCI} and \textit{30scenes}), etc.  
6 sampling patterns neighboring to the optimized one by our strategy were used for comparisons, as illustrated in the bottom row of Fig. \ref{fig_pattern_content_img}.
The PSNR of reconstructed LFs from inputs with different sampling patterns on LFs with different scene content is plotted in Fig. \ref{fig_pattern_content_psnr}.
It can be seen that although the PSNR values of reconstructed LFs present different trends when the sampling patterns change, the highest PSNR values have been achieved with the same sampling pattern by our method for most cases (9 out of 12).
In addition, our selected sampling pattern can achieve a comparable PSNR value to the highest one even when it is not optimal. Although the selected images cannot cover all scenarios, our experiment shows that the proposed optimization strategy is generally applicable in most of the cases we have experimented with.

\par
\textbf{$2)$ The effectiveness of the disparity estimation module.}

\par
In our approach, the disparity maps are estimated by constructing PSVs, which are fed into the subsequent network. 
Alternative ways include applying convolutional layers to the input SAIs straightly, or abstracting hand-craft features from PSVs as the input of a network \cite{lfrec2016kalantari}.
To validate the advantages of our disparity estimation module, 
we visually compared the by-product disparity maps estimated by these three manners.
Note that by training the network of Kalantari \textit{et al.} \cite{lfrec2016kalantari} using codes provided by the authors, the estimated disparity maps for the \textit{HCI} dataset are nearly all zeros. We believe the reason is that the only objective of the network is to optimize the final reconstruction by applying a loss function to the last refinement module. For LF datasets with large disparities, such a loss function can not efficiently back-propagate to the disparity values via warping operators. Therefore, we modified their source code, and re-trained their network by adding an intermediate supervision for the warped images using ground-truth targets, denoted as Kalantari \textit{et al.} \cite{lfrec2016kalantari} (inter).
Then the estimated disparity maps become reasonable. In addition, the average PSNR value of its final reconstructions on \textit{HCI} dataset is also improved by around 0.3 db.
As shown in Fig. \ref{fig_disp_inter},
it can be observed that our method produces disparity maps with much fewer error in both background and occlusion boundaries.

\textbf{$3)$ The effectiveness of the blending strategy.}

\par
The blending strategy in our approach is designed to address the occlusion issues during the fusion of the images warped from different input SAIs.
To validate the effectiveness of the proposed blending strategy, 
the intermediate results before and after blending are visualized in Fig. \ref{fig_blend}.
It can be observed that the errors around occlusion boundaries in
the intermediate images warped from different source SAIs 
are closely related to the location of the source SAIs, and appear in different positions. 
The learned confidence maps are able to indicate these error areas in each warped image, and provide guidance for the fusion of the warped images.
Blending over the guidance of confidence maps helps to remove these errors, while the correct regions of each warped image are preserved.

Moreover, to demonstrate the advantage of the proposed blending strategy, we quantitatively compared the blended results by our method and the method used in \cite{lfrec2016kalantari}. First, we removed the refinement module from our model, such that the remaining view synthesis network consists of disparity estimation, warping and the confidence-based blending. We denote this model as \textit{Ours\_conf\_blend}. Then, we replaced the confidence-based blending with the blending strategy used in [22], i.e., using convolutional layers to directly combine the warped image. This new model, denoted as \textit{Ours\_cnn\_blend}, was trained using the same datasets as ours. In this way, the only difference between these two models are the blending mechanisms. We compared their reconstruction quality, and the results are listed in Table \ref{table_blend}, where it can be seen that \textit{Ours\_conf\_blend} achieves higher PSNR/SSIM values than \textit{Ours\_cnn\_blend}, validating the advantage of our confidence-based blend strategy.

        \begin{table}[!t]
        \renewcommand{\arraystretch}{1.3}
        \caption{Effectiveness verification of the confidence-based fusion compared with blending using convolutional layers in \cite{lfrec2016kalantari}. The PSNR/SSIM values are provided for comparisons. $4(a)$ and $4(f)$ are two sampling patterns for the task $4\rightarrow 7\times7$ depicted in Fig. \ref{fig_pattern}.}
        \label{table_blend}
        \centering
        \resizebox{\linewidth}{!}{
        \begin{tabular}{c|c c | c c}
        \toprule[1.5pt]
        Test set  & \textit{Ours\_cnn\_blend}  & \textit{Ours\_conf\_blend}  & \textit{Ours\_cnn\_blend} &  \textit{Ours\_conf\_blend}  \\
        \midrule[1pt]
         & \multicolumn{2}{c|}{$4(a) \rightarrow 7\times7$} & \multicolumn{2}{c}{$4(f) \rightarrow 7\times7$} \\
        \textit{30scenes} & 39.78/0.978  & \textbf{40.12}/\textbf{0.979} & 41.17/0.983 & \textbf{41.57}/\textbf{0.983} \\
        \textit{Occlusions} & 35.30/0.963 & \textbf{35.42}/\textbf{0.963} & 37.34/0.974 & \textbf{38.14}/\textbf{0.976}\\
        \textit{Reflective} & 35.63/0.938 & \textbf{36.45}/\textbf{0.941} & 38.59/0.955 & \textbf{39.24}/\textbf{0.957} \\
        \bottomrule[1.5pt]
        \end{tabular}
        }
        \end{table}

\textbf{$4)$ The effectiveness of the refinement module.}

        \begin{figure*}[!t]
        \centering
        \includegraphics[width=0.8\linewidth]{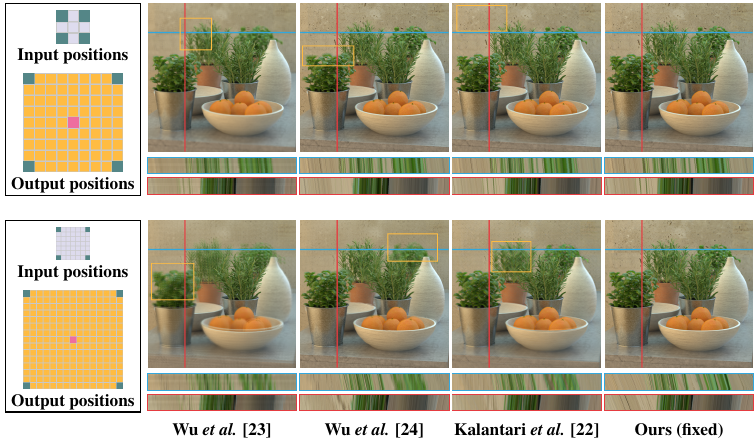}
        \caption{Visual comparisons on LF reconstruction with flexible output angular resolution. We present the results of $9\times9$ reconstruction from 4 corner SAIs of a $3\times3$ sampling grid (top), and the results of $15\times15$ reconstruction from 4 corner SAIs of a $7\times7$ sampling grid (bottom). The center SAI of the LF images reconstructed from different algorithms are presented. Horizontal and vertical EPIs corresponding to  the colored lines are shown below the center SAI, and regions with obvious artifacts or blurring are highlighted with yellow boxes. It is recommended to view this figure by zooming in.}
        \label{fig_denser}
        \end{figure*}

\par
To demonstrate the effectiveness of the refinement module, we quantitatively compared the quality of the LF images generated by our method without the refinement module and the LF images by our method with all modules, and Table \ref{table_refine} lists the results. 
It can be seen that the refinement provides around 1 dB PSNR improvement, which indicates that the refinement module can efficiently exploit the complementary information between the synthesized SAIs and improves the intermediate LF images.

        \begin{figure*}[!t]
        \centering
        \includegraphics[width=\linewidth]{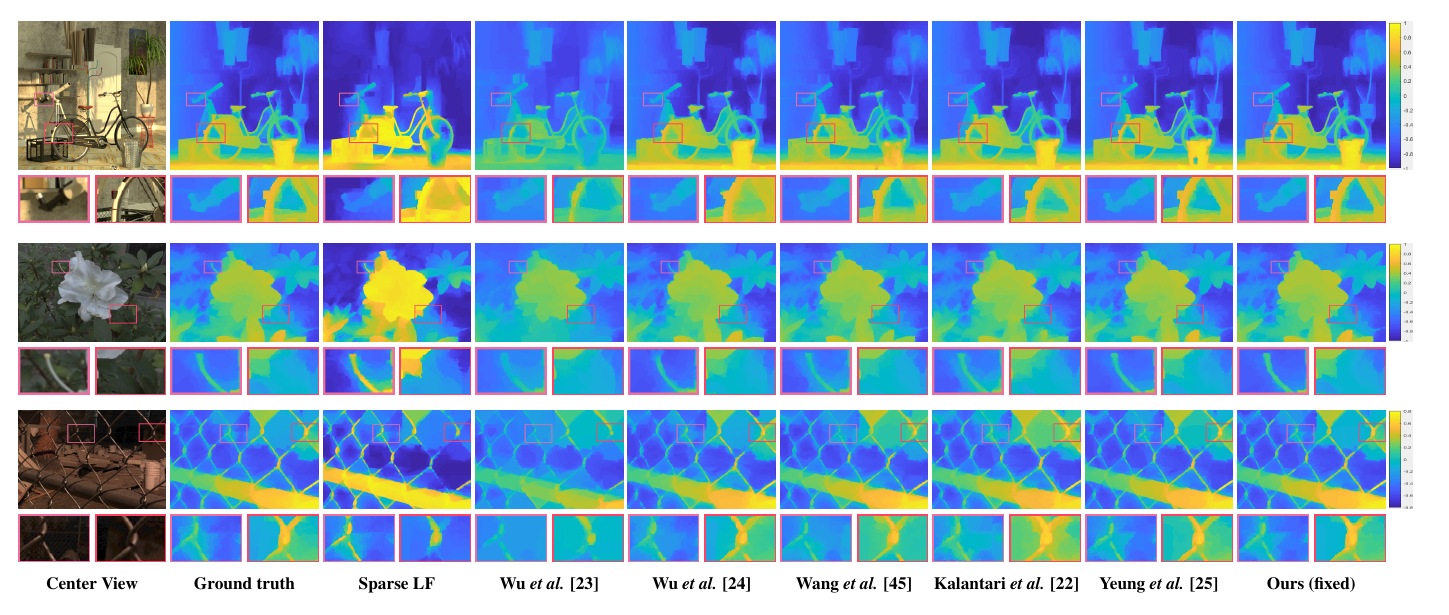}
        \caption{Visual comparisons of the depth estimation results (as the depth is inverse proportional to the disparity, we do not make a distinction between them).  The center SAIs of the LF images, the disparity maps estimated from the ground-truth densely-sampled LFs, the  sparsely-sampled  LFs, the reconstructed densely-sampled  LFs by different algorithms are presented from left to right. It is recommended to view this figure by zooming in.}
        \label{fig_disp}
        \end{figure*}

       \begin{table*}[!t]
        \renewcommand{\arraystretch}{1.3}
        \caption{Quantitative comparisons (100$\times$ MSE) of the depth estimated from the ground-truth densely-sampled light fields, the sparsely-sampled light fields, the reconstructed densly-sampled light fields by different methods. The three numbers from left to right are the results of \cite{lfdepth2018chen}$/$\cite{lfdepth2016spo}$/$\cite{lfdepth2015wang}. The lower, the better. The best and second best results of different reconstruction methods under an identical depth estimation algorithm are highlighted in \textcolor{red}{red} and \textcolor{blue}{blue}, respectively.
        }
        \label{table_disp_enhance}
        \centering
        \resizebox{\textwidth}{!}{
        \begin{tabular}{l| l l l l l l l l }
        \toprule[1.5pt]
        Reconstruction Method & \textit{Buddha} & \textit{Buddha2} & \textit{StillLife} & \textit{Pupillon} & \textit{Monasroom}\\
        \midrule[1pt]
        Ground-truth LF & \textcolor{blue}{0.79}/1.10/\textcolor{red}{0.73} & \textcolor{blue}{0.54}/2.22/0.96 &  \textcolor{red}{4.56}/\textcolor{red}{1.88}/\textcolor{red}{2.73} &  1.43/\textcolor{blue}{1.31}/0.78 & \textcolor{blue}{0.50}/0.64/0.55 \\
        Sparse LF & 0.85/1.84/44.51 & 0.64/3.43/160.52 & \textcolor{blue}{4.88}/4.91/217.79 & \textcolor{blue}{1.29}/6.83/24.10 & \textcolor{red}{0.49}/1.07/29.50 \\
        Wu \textit{et al.} \cite{lfrec2018wu:blur} & 6.06/45.88/8.64 & 2.62/36.81/4.24 & 100.75/183.47/199.88 & 3.89/34.14/9.99 & 1.28/10.52/1.35 \\
        Wu \textit{et al.} \cite{lfrec2019wu:shear} & 1.03/1.32/1.29 & 0.69/0.57/0.76 & 19.33/69.14/61.68 & 1.67/3.22/31.82 & 0.66/1.04/0.90 \\
        Wang \textit{et al.} \cite{lfrec2018p4d} & 1.29/0.96/1.35 & 0.67/0.59/0.67 & 47.25/68.04/81.24 & 1.74/2.93/1.21 & 0.90/1.18/0.80 \\
        Kalantari \textit{et al.} \cite{lfrec2016kalantari}  & 1.08/\textcolor{blue}{0.85}/0.96  & 0.61/\textcolor{blue}{0.52}/\textcolor{blue}{0.56} & 18.88/70.50/60.71 &  1.95/2.50/0.96 &  0.61/1.16/\textcolor{blue}{0.46} \\
        Yeung \textit{et al.} \cite{lfrec2018Yeung} & 0.86/\textcolor{red}{0.61}/0.81  & 0.58/\textcolor{red}{0.51}/0.67  &  35.24/102.05/100.82 & 1.66/1.98/\textcolor{blue}{0.64} & \textcolor{blue}{0.50}/\textcolor{red}{0.49}/\textcolor{red}{0.39}\\
        Ours (fixed) & \textcolor{red}{0.78}/0.97/\textcolor{blue}{0.79} & \textcolor{red}{0.48}/0.61/\textcolor{red}{0.45} & 5.35/\textcolor{blue}{3.94}/\textcolor{blue}{4.52} & \textcolor{red}{1.05}/\textcolor{red}{1.06}/\textcolor{red}{0.57} & \textcolor{red}{0.49}/\textcolor{blue}{0.57}/\textcolor{red}{0.39}\\
        \bottomrule[1.5pt]
        \end{tabular}
        }
        \end{table*}

        \begin{table*}[!t]
        \renewcommand{\arraystretch}{1.3}
        \caption{Quantitative comparisons (Bad Pixel Ratios with threshold 0.07) of the depth estimated from the ground-truth densely-sampled light fields, the sparsely-sampled light fields, the reconstructed densly-sampled light fields by different methods. The three numbers from left to right are the results of  \cite{lfdepth2018chen}$/$\cite{lfdepth2016spo}$/$\cite{lfdepth2015wang}. The lower, the better. The best and second best results of different reconstruction methods under an identical depth estimation algorithm are highlighted in red and blue, respectively.
        }
        \label{table_disp_enhance_bp7}
        \centering
        \resizebox{\textwidth}{!}{
        \begin{tabular}{l| l l l l l l l l }
        \toprule[1.5pt]
        Reconstruction Method & \textit{Buddha} & \textit{Buddha2} & \textit{StillLife} & \textit{Pupillon} & \textit{Monasroom}\\
        \midrule[1pt]
        Ground-truth LF & \textcolor{red}{11.90}/\textcolor{blue}{3.95}/5.50 & 17.66/23.08/18.79 &  \textcolor{red}{39.73}/\textcolor{red}{11.70}/\textcolor{red}{14.56} &  26.56/20.39/24.53 & \textcolor{blue}{12.61}/8.49/9.50 \\
        Sparse LF & 12.51/\textcolor{red}{3.70}/93.03 & 18.75/18.44/91.04 & \textcolor{blue}{42.85}/\textcolor{blue}{13.07}/96.50 & \textcolor{blue}{25.90}/15.28/92.50 & \textcolor{red}{12.07}/7.72/94.39 \\
        Wu \textit{et al.} \cite{lfrec2018wu:blur} & 66.81/22.65/67.28 & 60.42/27.34/42.59 & 93.46/65.78/90.55 & 60.51/30.89/67.26 & 51.03/16.71/38.54 \\
        Wu \textit{et al.} \cite{lfrec2019wu:shear} & 13.13/5.67/\textcolor{blue}{5.47} & 16.79/\textcolor{blue}{11.19}/\textcolor{blue}{6.90} & 81.30/39.49/78.78 & 33.03/\textcolor{blue}{12.23}/\textcolor{blue}{9.22} & 15.34/8.08/7.48 \\
        Wang \textit{et al.} \cite{lfrec2018p4d} & 22.55/9.59/16.76 & 20.80/18.06/12.45 & 84.62/43.12/81.87 & 38.19/13.64/18.26 & 23.61/9.85/11.06 \\
        Kalantari \textit{et al.} \cite{lfrec2016kalantari}  & 17.36/8.33/7.14  & \textcolor{blue}{16.08}/13.89/9.13 & 83.33/40.58/87.85 &  34.08/14.76/12.24  &  16.93/9.53/8.99 \\
        Yeung \textit{et al.} \cite{lfrec2018Yeung} & 14.95/7.53/5.48  & 16.91/14.75/10.10  &  85.78/47.31/86.62 & 31.77/13.01/10.67 & 13.46/\textcolor{blue}{7.57}/\textcolor{blue}{6.79}\\
        Ours (fixed) & \textcolor{blue}{12.09}/4.96/\textcolor{red}{4.87} & \textcolor{red}{12.95}/\textcolor{red}{10.32}/\textcolor{red}{6.86} & 48.06/22.53/\textcolor{blue}{34.12} & \textcolor{red}{22.54}/\textcolor{red}{11.06}/\textcolor{red}{7.20} & 12.24/\textcolor{red}{6.98}/\textcolor{red}{5.94}\\
        \bottomrule[1.5pt]
        \end{tabular}
        }
        \end{table*}

\section{Applications}
\label{sec:application}

In this section, we will discuss two applications, which will benefit from our accurate, flexible and efficient method for the reconstruction of densely-sampled LFs.

\subsection{Image-based rendering (IBR)}
IBR aims at generating novel views from a set of captured images.
Comprehensive review on IBR can be found in \cite{viewsyn2004survey}.
Among IBR techniques, LF rendering is attractive as novel views can be generated by straightforward interpolation without the need of any geometric information such that real-time rendering can be achieved. To produce novel views without ghosting artifacts, LF rendering requires the LF to be densely sampled, with disparities between neighboring views to be less than 1 pixel \cite{chai2000plenopticsampling}.
Therefore,  for a sparsely-sampled  LF that does not meet the sampling requirement, our method can reconstruct a densely-sampled  LF with  desired angular resolution to enable  subsequent LF rendering. More generally, as our method is capable of generating novel views at arbitrary viewpoints from a set of sparsely-sampled SAIs, it can realize IBR directly.

\par
To validate the effectiveness of our approach on the IBR application, we performed comparisons of dense reconstruction under different sampling baselines for different output angular resolution. Specifically, we compared the performance of different algorithms when reconstructing $9\times9$ densely-sampled  LFs from $2\times2$ corner SAIs sampled at a $3\times3$ grid, and reconstructing $15\times15$ densely-sampled  LFs from $2\times2$ corner SAIs sampled at a $7\times7$ grid on \textit{HCI} dataset.
As the ground-truth images are unavailable, we visually compared the center SAIs of reconstructed LF images.
Moreover, to compare the ability of preserving the LF parallax structure, horizontal and vertical EPIs are presented.
Fig. \ref{fig_denser} shows the results, and it can be observed that our method can produce novel SAIs with sharp textures and construct EPIs with clear linear structures, even when the input sampling baselines are relatively large.

\subsection{Depth estimation enhancement}
\label{sec:depth_enhanc}
The value of an LF image lies in the implicitly encoded scene geometry information.
By finding correspondences in different SAIs, depth maps can be estimated from the LF images.
A densely-sampled  LF leads to more accurate and more robust depth inference, as matching points can be detected more easily and occlusion problems can be alleviated by multiple viewpoints.
Therefore, the proposed method can be used to enhance LF depth estimation.

\par
Here, we present the depth maps estimated from sparsely-sampled $3\times3$ LF images as well as those estimated from densely-sampled $7\times7$ LF images reconstructed by different algorithms.
The state-of-the-art depth estimation algorithm \cite{lfdepth2018chen} was applied, and Fig. \ref{fig_disp} shows the results.
It can be observed that the reconstructed densely-sampled  LFs enable better estimations than sparsely-sampled LF ones, and
the depth maps from our method are more accurate than those from others, especially in the regions including fine details and occluded boundaries. 
Additionally, the high accuracy of estimated depth maps further validates the advantage of our method on  preserving the LF parallax structure.

\par
Moreover, we provided quantitative comparisons of the depth maps estimated from different reconstructions.
For robust and reliable evaluation, 3 widely-used and robust  depth estimation methods, i.e., \cite{lfdepth2018chen, lfdepth2016spo, lfdepth2015wang}, were used to avoid possible errors or possible adaptation to the reconstruction methods.
Mean square error (MSE) and  Bad Pixel Ratio (BPR) between the estimated depth map and its ground-truth were used to measure the accuracy.
BPR measures the percentage of pixels with an error large than the threshold in the estimated depth map.
Tables \ref{table_disp_enhance} and \ref{table_disp_enhance_bp7} list the results, where it can be seen that when evaluated with different depth estimation algorithms, the MSE and BPR of
\textit{Ours (fixed)} are the lowest or second lowest under most cases, compared with other methods, especially on the LF image with a large disparity, i.e., \textit{StillLife}.
Particularly, the MSE values of \textit{Ours (fixed)} are even lower than those of the depth maps estimated from the ground-truth densely-sampled LFs in some cases. The reason could be twofold:
no method can guarantee perfectly accurate estimations, and sometimes the adopted depth estimation methods adapt to the reconstructed LFs by our method better; and much noise is present in the raw LF images \cite{lfdataset2016hci}, while the noise might be suppressed by our reconstruction algorithm to some extent.

\section{Conclusion  and Future Work}
\label{sec:conclusion}
We have presented a novel learning-based algorithm for the reconstruction of densely-sampled LFs from sparsely-sampled ones.
Owing to the deep, effective and comprehensive modeling of the unique LF parallax structure, including the geometry-based SAI synthesis based on position-aware PSVs, the adaptive blending strategy and the efficient LF refinement network, our method breaks the obstacle in an arbitrary sampling pattern and sparse sampling, not only achieving over 4 dB improvement on synthetic data and 1 dB improvement on real-world data, but also preserving the valuable LF parallax structure better, compared with state-of-the-art methods.
Besides, we proposed a simple yet effective algorithm  to optimize the sparse sampling pattern for better reconstruction quality.
Last but not least, the potential of our method on improving  subsequent LF-based applications has been validated and discussed.

During the sampling pattern optimization, we have built a scene content-independent strategy, which only considers the overall distance between the novel views and the sampled ones and the distribution divergence of the sampling.
In fact, the optimal sampling pattern should vary with the scene content, such as the geometry complexity and textual information.
In our future work, we plan to predict the scene content-dependent optimized sampling pattern via a CNN trained with the ground-truth optimal sampling patterns that can be obtained via an exhaustive search.

Second, we evaluated the quality of reconstructed LFs through the average of SAI-wise
PSNR/SSIM and estimated depth maps. However, these metrics can only evaluate the LF image on the spatial dimension, or evaluate the angular consistency indirectly. It is thus highly desirable to build a standard and effective metric for evaluating the quality of 4-D LFs directly.
Researchers from the image quality assessment field have started paying attention to this issue
\cite{LFmetric2017mpi,LFmetric2019depth,LFmetric2020tianyu}.

Another interesting line of future work is exploring the potential of the proposed framework on LF data compression.
The huge data size of LF images poses great challenges to both data storage and transmission. In \cite{lfcompression2018hou}, an LF image is partitioned into key SAIs and non-key SAIs, and non-key SAIs are compensated by the reconstruction from key SAIs.
Only the key SAIs and residual of non-key SAIs are encoded.
Our framework adapting to flexible inputs can be naturally utilized to optimize the combination of key SAIs so that the reconstruction quality of non-key SAIs can be improved using the same number of key SAIs, and likewise the compression performance.
Moreover, our experimental results have demonstrated that using the optimized sampling patterns, the number of key SAIs can be reduced without penalizing the reconstruction performance, which means the encoding bits of key SAIs can be saved.
In the future, we will comprehensively study how the sampling pattern and the number of input views affect the compression performance, and experimentally verify the application of the proposed framework on LF compression.

\section*{Acknowledgement}

We thank the authors of \cite{lfrec2018p4d} for sharing their source codes.

\normalem
\bibliographystyle{IEEEtran}
\bibliography{IEEEabrv,./references}

\end{document}